\documentclass[aps,prb,twocolumn,groupedaddress]{revtex4-1}
\usepackage{graphicx}
\usepackage{dcolumn}
\usepackage{bm}
\usepackage{amsmath}
\usepackage{amssymb}
\usepackage{color}

\begin{document}

\title{Curvature-sensing and generation by membrane proteins: a review}

\author{Hiroshi Noguchi}
\email[]{noguchi@issp.u-tokyo.ac.jp}
\affiliation{Institute for Solid State Physics, University of Tokyo, Kashiwa, Chiba 277-8581, Japan}

\begin{abstract}
Membrane proteins are crucial in regulating biomembrane shapes and controlling the dynamic changes in membrane morphology during essential cellular processes. These proteins can localize to regions with their preferred curvatures (curvature sensing) and induce localized membrane curvature.
Thus, this review describes the recent theoretical development in membrane remodeling performed by membrane proteins.
The mean-field theories of protein binding and
the resulting membrane deformations are reviewed. The effects of hydrophobic insertions on the area-difference elasticity energy
and that of intrinsically disordered protein domains on the membrane bending energy are discussed.
For the crescent-shaped proteins, such as Bin/Amphiphysin/Rvs superfamily proteins,
anisotropic protein bending energy and orientation-dependent excluded volume significantly contribute to
curvature sensing and generation.
Moreover, simulation studies of membrane deformations caused by protein binding
are reviewed, including domain formation, budding, and tubulation.
\end{abstract}
\maketitle

\section{Introduction}

Cell membranes and organelles exhibit a variety of shapes.
Various types of proteins are known to control dynamical changes in membrane morphology during essential cellular processes, such as
endocytosis, exocytosis, vesicle transport, mitosis, and cell locomotion.\cite{mcma05,zimm06,suet14,kaks18,beth18,svit18,lutk12} 
During {\textit{in vitro}} experiments, protein binding has been observed to induce membrane budding and tubulation.
Additionally, membrane proteins can localize to membrane regions of specific curvature.
These two phenomena are referred to as curvature generation and curvature sensing, respectively.
This review focuses on the theoretical studies of protein behaviors with emphasis on thermal equilibrium and relaxation to the equilibrium.
Non-equilibrium membrane dynamics, such as non-thermal fluctuations\cite{turl19,mann01,lenz03,nogu21} and wave propagations,\cite{wu18,taka22,pele11,tame20,tame21,tame22,nogu23a} are covered in our recent review.~\cite{nogu24c}
Moreover, because  membrane simulation models and methods have previously been reviewed in Refs.~\citenum{muel06,vent06,nogu09,shin12,peze21,marr23,dasa24}, this review primarily describes the mean-field theory and presents relevant simulation results without delving into detailed simulation methodologies.

Section~\ref{sec:lip} provides an overview of the bending energy of lipid membranes and their morphology in the absence of proteins.
Section~\ref{sec:sens} discusses curvature-sensing.
Certain proteins exhibit laterally isotropic shapes in a membrane and bend the membrane isotropically.
Section~\ref{sec:isosens} presents the theoretical aspects of isotropic proteins, and
section~\ref{sec:idp} explores how intrinsically disordered protein (IDP) domains influence membrane bending properties.
Section~\ref{sec:anisens} addresses the behavior of anisotropic proteins.
Crescent-shaped proteins, such as Bin/Amphiphysin/Rvs (BAR) superfamily proteins, induce membrane bending along their major protein axes.
Section~\ref{sec:tether} examines protein binding to tethered vesicles and presents the estimation of protein bending properties.
Section~\ref{sec:gen} focuses on curvature generation, with
Sections~\ref{sec:isogen} and \ref{sec:anigen} reviewing membrane deformations induced by the  isotropic and anisotropic proteins, respectively.
Section~\ref{sec:nano} discusses the membrane deformation by the adhesion of colloidal nanoparticles.
Finally, Section~\ref{sec:sum} provides a summary and outlook.

\section{Lipid Membranes}\label{sec:lip}

\begin{figure}[t]
\includegraphics[]{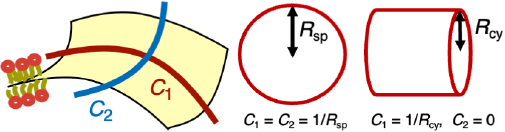}
\caption{
Schematic of lipid membranes.
A membrane locally bends with two principal curvatures $C_1$ and $C_2$.
A spherical membrane has $C_1=C_2=1/R_{\mathrm{sp}}$ ($H=1/R_{\mathrm{sp}}$ and $K=1/R_{\mathrm{sp}}^2$),
while a cylindrical membrane has $C_1=1/R_{\mathrm{cy}}$ and $C_2=0$ ($H=1/2R_{\mathrm{cy}}$ and $K=0$).
}
\label{fig:cartmb}
\end{figure}

In a fluid phase, lipid membranes are laterally isotropic, and 
their bending energy can be expressed using the second-order expansion of the membrane curvature, 
known as the Canham--Helfrich model.\cite{canh70,helf73} 
\begin{equation}
F_{\mathrm{cv0}} = \int \Big[ \frac{\kappa_{\mathrm{d}}}{2}(2H - C_{\mathrm{mb}})^2 + \bar{\kappa}_{\mathrm{d}}K \Big] \mathrm{d}A,
\label{eq:fmb}
\end{equation}
where  $A$ represents the membrane area.
The membrane mean and Gaussian curvatures are defined as
$H= (C_1+C_2)/2$ and $K=C_1C_2$, respectively, where
 $C_1$  and $C_2$ represent the principal curvatures (see Fig.~\ref{fig:cartmb}).
The coefficients $\kappa_{\mathrm{d}}$ and $\bar{\kappa}_{\mathrm{d}}$ denote the bending rigidity and saddle-splay modulus 
(also referred to as the Gaussian curvature modulus),
respectively. The parameter $C_{\mathrm{mb}}$ denotes the spontaneous curvature. 
Note that the spontaneous curvature is often expressed as $H_{\mathrm{mb}}=C_{\mathrm{mb}}/2$, which is particularly useful in the analysis of  spherical membranes, while $C_{\mathrm{mb}}$ is useful for cylindrical membranes.
For lipid bilayers with symmetric leaflets,
the membrane has zero  spontaneous curvature ($C_{\mathrm{mb}}=0$). 
The last term in Eq.~(\ref{eq:fmb}) can be neglected when considering the shape transformation of vesicles
with a fixed topology,
owing to the Gauss--Bonnet theorem, $\oint C_1C_2 \mathrm{d}A= 4\pi (1-g)$, where $g$ represents the genus of the vesicle.
Lipid membranes typically exhibit a bending rigidity in the range of $\kappa_{\mathrm{d}} = 10$--$100k_{\mathrm{B}}T$\cite{kara23,dimo14,mars06,rawi00}
and $\bar{\kappa}_{\mathrm{d}}/\kappa_{\mathrm{d}} \simeq -1$.\cite{hu12},
where $k_{\mathrm{B}}T$ is the thermal energy.
In this review, we use $\kappa_{\mathrm{d}} = -\bar{\kappa}_{\mathrm{d}} = 20k_{\mathrm{B}}T$ and $C_{\mathrm{mb}}=0$, unless otherwise specified.

In lipid membranes,
the traverse movement of phospholipids between the two leaflets, known as flip--flop,
occurs at an extremely slow rate, with half-lives ranging from hours to days.\cite{korn71}
In contrast, amphiphilic molecules with small hydrophilic head groups, such as 
cholesterols, exhibit significantly faster flip--flop dynamics, occurring within seconds to minutes.\cite{cont09,hami03,stec02}
In living cells, proteins facilitate flip--flop.
Flippase and floppase proteins actively transport specific lipids from the outer to the inner leaflets (flip) or in the opposite direction (flop), respectively, through ATP hydrolysis, 
leading to an asymmetric lipid distribution. Conversely,
scramblases mediate the bidirectional translocation of lipids, allowing the bilayer to relax toward a thermal-equilibrium lipid distribution.\cite{cont09}
As a result, the number of lipids in each leaflet remains constant over typical experimental timescales,
although it can relax with the addition of cholesterols,\cite{bruc09,miet19}  ultra-long-chain fatty acids,\cite{kawa20,kawa22} and scramblases.
Consequently, the area difference $\Delta A= 2h\oint H \mathrm{d}A$ of the two leaflets in a liposome may differ from the 
lipid-preferred area difference $\Delta A_0=(N_{\mathrm{out}}-N_{\mathrm {in}})a_{\mathrm {lip}}$, where $N_{\mathrm{out}}$ and $N_{\mathrm{in}}$
represent the numbers of lipids in the outer and inner leaflets, respectively,
 $a_{\mathrm {lip}}$ is the area per lipid, and
 $h \simeq 2$\,nm denotes the distance between the centers of the two leaflets.
In the area difference elasticity (ADE) model,\cite{seif97,svet09,svet89} 
the energy associated with the mismatch $\Delta A-\Delta A_0$ is accounted for by a harmonic potential:
\begin{eqnarray}
F_{\mathrm{ade}} &=&  \frac{\pi k_{\mathrm{ade}}}{2Ah^2}(\Delta A - \Delta A_0)^2 \label{eq:ade0} \\
&=&  \frac{k_{\mathrm{ade}}}{2}(m - m_0)^2  \label{eq:ade1} \\
 &=&  8\pi k_{\mathrm{r}}(\Delta a - \Delta a_0)^2,
\label{eq:ade2}
\end{eqnarray}
where the coefficient $k_{\mathrm{r}}= \pi k_{\mathrm{ade}}$.
In Eqs.~(\ref{eq:ade1}) and (\ref{eq:ade2}), 
the area differences are normalized as $m= \Delta A/2hR_{\mathrm{A}}$ and $\Delta a= \Delta A/8\pi hR_{\mathrm{A}}$,
where the lengths are normalized using the vesicle surface area as $R_{\mathrm{A}}=(A/4\pi)^{1/2}$.
These two formulations were used in Ref.~\citenum{seif97} and Refs.~\citenum{svet09,svet89}, respectively.
For typical lipid membranes,
 $k_{\mathrm{ade}} \simeq \kappa_{\mathrm{d}}$ was estimated.\cite{saka12}

\begin{figure}[t]
\includegraphics[]{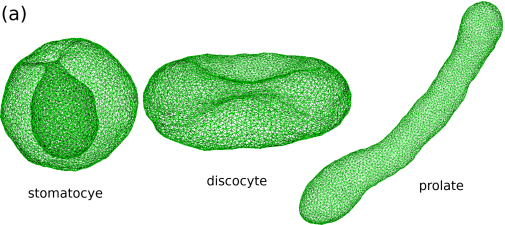}
\includegraphics[]{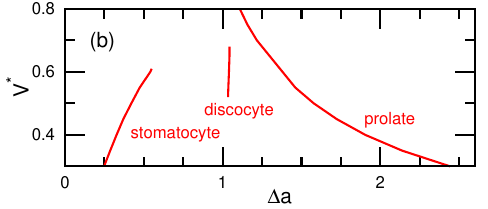}
\caption{
Stable and meta-stable shapes of vesicles in the Canham--Helfrich model (Eq.~(\ref{eq:fmb})) with $C_{\mathrm{mb}}=0$.\cite{nogu05,nogu15a}
(a) Snapshots obtained by dynamically-triangulated MC simulations.
Stomatocyte at $V^*=0.5$, discocyte at $V^*=0.6$, and prolate at $V^*=0.5$.
(b) Area difference $\Delta a$ of (meta-) stable shapes.
 Adapted from Ref.~\citenum{nogu15a} with permission from the Royal Society of Chemistry (2015).
}
\label{fig:mbves}
\end{figure}

Because the critical micelle concentration (CMC) of lipids 
is extremely low,\cite{tanf80}
 the number of lipid molecules within a vesicle remains essentially constant over typical experimental
timescales. Additionally, the internal volume is maintained nearly constant
due to osmotic pressure, since
water molecules can slowly permeate the lipid bilayer, whereas
the penetration of ions or macromolecules is negligible.
Under the constraints of 
a constant volume $V$ and constant surface area $A$ at $C_{\mathrm{mb}}=0$, 
the global energy minimum of $F_{\mathrm{cv0}}$
corresponds to different vesicle shapes depending on the reduced volume
 $V^*= 3V/(4\pi R_{\mathrm{A}}^3)$.
In the mechanical (force) viewpoint, the stress caused by the bending energy
is balanced by the surface tension ($\gamma$) and the osmotic pressure difference between the inner and outer solutions.
For vesicles with genus $g=0$,
stomatocyte, discocyte, and prolate shapes
achieve global energy minima within the ranges
$0<V^* \lesssim 0.59$, $0.59 \lesssim V^* \lesssim 0.65$, 
and $0.65 \lesssim V^* < 1$, respectively.\cite{lipo95,seif97,seif91}
These three shapes can coexist as (meta-)stable states at $V^* \simeq 0.6$,\cite{nogu05}
and the prolate shape can persist as a meta-stable state even at $V^* \lesssim 0.6$, as illustrated in Fig.~\ref{fig:mbves}.\cite{nogu15a,bahr17} 
Note that red blood cells have a discocyte shape with $V^* \simeq 0.6$ in the physiological condition,
and their membranes have shear elasticity due to the cytoskeletons underneath the membranes.\cite{lim08}
When the ADE energy is included,
additionally, branched tubular vesicles
and budding (where spherical buds form on the outside of a spherical vesicle)\cite{seif97,zihe05}
emerge alongside the stomatocyte, discocyte, and prolate shapes.
Notably, experimental observations have been well reproduced by this theoretical model.\cite{saka12} 
Furthermore, rapid changes in $\Delta A_0$ induced by chemical reactions and other factors
can lead to the protrusion of bilayer sheets, reducing the area difference.\cite{naka18,sree22}

\begin{figure}[t]
\includegraphics[]{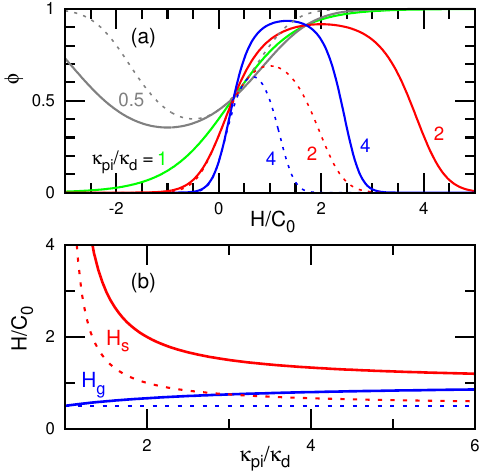}
\caption{
Curvature sensing of isotropic proteins at $C_0^2 a_{\mathrm{p}} = 0.04$ (e.g., $C_0= 0.02$\,nm$^{-1}$ for $a_{\mathrm{p}}=100$\,nm$^2$) and  
$\bar{\kappa}_{\mathrm{pi}}/\kappa_{\mathrm{pi}}=\bar{\kappa}_{\mathrm{d}}/\kappa_{\mathrm{d}}= -1$.\cite{nogu21a,nogu22a} 
(a) Protein density $\phi$ as a function of the local mean curvature $H$ 
at $\kappa_{\mathrm{pi}}/\kappa_{\mathrm{d}}=0.5$, $1$, $2$, and $4$ for $\mu=0$.
(b) Sensing curvature $H_{\mathrm{s}}$ and the maximum generation curvature $H_{\mathrm{g}}$
as a function of bending rigidity ratio $\kappa_{\mathrm{pi}}/\kappa_{\mathrm{d}}$.
$H_{\mathrm{s}}$ and  $H_{\mathrm{g}}$ are given by Eqs.~(\ref{eq:hs}) and (\ref{eq:hg}) with $\phi=1$, respectively.
The solid and dashed lines represent the data for spherical and
cylindrical membranes, respectively.
}
\label{fig:seniso}
\end{figure}

\section{Curvature Sensing}\label{sec:sens}

Peripheral and transmembrane proteins tend to accumulate in membrane regions that match their preferred curvatures.
The surface densities of peripheral proteins are governed by the balance between 
the protein binding and unbinding processes in thermal equilibrium.
In contrast, in typical \textit{in vitro} experiments,
 the total number of transmembrane proteins within a vesicle remains fixed.
These scenarios correspond to grand canonical and canonical ensembles in the membrane, respectively.
While the choice between these two conditions does not affect average properties, such as surface protein density and alignment,
it influences kinetics and fluctuations (the second derivatives of free energy).

\subsection{Theory of Isotropic Proteins}\label{sec:isosens}

First, we discuss the curvature-sensing phenomenon of proteins with a laterally isotropic shape.
The insertion of a hydrophobic $\alpha$-helix and the anchoring of IDP domains
do not exhibit a preferred bending direction.
Moreover, proteins or protein assemblies possessing threefold, fivefold, or higher rotational symmetry 
exhibit laterally isotropic bending energy, when their asymmetric deformations are negligible.\cite{nogu24}
Several types of ion channels~\cite{tray10,syrj21} and G-protein coupled receptors (GPCRs)~\cite{erns14,naga21,venk13,shib18},
 have rotationally symmetric structures.
For instance, the trimer and pentamer of microbial rhodopsins exhibit threefold and fivefold symmetry, respectively~\cite{shib18}.
Furthermore, certain peripheral proteins, such as a clathrin monomer~\cite{hurl10} and annexin A5 trimer~\cite{gerk05,olin00}, 
also possess threefold symmetry.

The presence of membrane-bound proteins can alter the membrane bending rigidity and spontaneous curvature relative to a bare (unbound) membrane.
The bending energy of a vesicle can be expressed as\cite{nogu21a}
\begin{eqnarray}\label{eq:Fcv0}
F_{\mathrm{cv1}} &=&  4\pi\bar{\kappa}_{\mathrm{d}}(1-g) + \int {\mathrm{d}}A \ \Big\{ 2\kappa_{\mathrm{d}}H^2(1-\phi)  \nonumber \\
&&  + \frac{\kappa_{\mathrm{pi}}}{2}(2H-C_0)^2\phi  
+ (\bar{\kappa}_{\mathrm{pi}}-\bar{\kappa}_{\mathrm{d}})K \phi  \Big\},
\end{eqnarray}
where 
$\kappa_{\mathrm{pi}}$, $\bar{\kappa}_{\mathrm{pi}}$, and $C_0=2H_0$ denote the bending rigidity, saddle-splay modulus, 
and spontaneous curvature of the protein-occupied membrane, respectively,
and $\phi$ represents the local protein density (area fraction, i.e., $\phi=1$ indicates complete coverage).
This formulation accounts for the bending energy induced by the protein-membrane interactions.
Additionally, inter-protein interactions --such as the steric effects arising from the brush region of IDP chains discussed in Section~\ref{sec:idp}--
 can further influence membrane rigidity and spontaneous curvature.
In that case, $\kappa_{\mathrm{pi}}$, $\bar{\kappa}_{\mathrm{pi}}$, and $C_0$ become functions of $\phi$.\cite{nogu21a}

At $\kappa_{\mathrm{pi}}>\kappa_{\mathrm{d}}$,
the bending energy can also be expressed as $F_{\mathrm{cv1}} = F_{\mathrm{cv0}} + F_{\mathrm{pi}}$
with
\begin{equation}\label{eq:Fcv1}
F_{\mathrm{pi}} = \int {\mathrm{d}}A \ 
    \big[ \frac{\kappa_{\mathrm{pa}}}{2}(2H-C_{\mathrm{0a}})^2 + \bar{\kappa}_{\mathrm{pa}}K \big]\phi,
\end{equation}
where   $\kappa_{\mathrm{pi}}=\kappa_{\mathrm{pa}}+\kappa_{\mathrm{d}}$, $\bar{\kappa}_{\mathrm{pi}}=\bar{\kappa}_{\mathrm{pa}}+\bar{\kappa}_{\mathrm{d}}$
and $C_0= [\kappa_{\mathrm{pa}}/(\kappa_{\mathrm{pa}}+\kappa_{\mathrm{d}})] C_{\mathrm{0a}}$.
This formulation is known as the curvature mismatch model, where
 $\kappa_{\mathrm{pa}}$ represents the additional bending rigidity by protein binding,
while $\kappa_{\mathrm{pi}}$ accounts for the combined rigidity of the protein and the underlying membrane.
The curvature mismatch model with $\bar{\kappa}_{\mathrm{pa}}=0$ was used in Refs.~\citenum{prev15,rosh17,tsai21}.

The total free energy $F$ of a vesicle consists of the bending energy $F_{\mathrm{cv1}}$, the inter-protein interaction energy, and the mixing entropy:
\begin{equation}\label{eq:F0}
F =  F_{\mathrm{cv1}} + \int {\mathrm{d}}A \ \Big\{ b \phi^2 +
  \frac{k_{\mathrm{B}}T}{a_{\mathrm{p}}}[\phi \ln(\phi) +  (1-\phi) \ln(1-\phi) ] \Big\},
\end{equation}
where 
 $a_{\mathrm{p}}$ denotes the area occupied by a single protein.
The second term in Eq.~(\ref{eq:F0}) represents the pairwise inter-protein interactions, where
$b<0$ and $b>0$ indicate attractive and repulsive interactions between bound proteins, respectively.
The third term in Eq.~(\ref{eq:F0}) accounts for the mixing entropy of the bound proteins.

The binding equilibrium of peripheral proteins is determined by minimizing $J=F-\mu N$, 
where $\mu$ is the binding chemical potential of the protein binding, 
and $N=\int \phi\ {\mathrm{d}}A/a_{\mathrm{p}}$ is the number of the bound proteins. 
Consequently, the local protein density $\phi$ is given by $\partial f/\partial \phi = \mu/a_{\mathrm{p}}$,
where $F = \int f\ {\mathrm{d}}A$.
When the inter-protein interactions are negligible ($b=0$),
 $\phi$ is expressed by a sigmoid function of $\mu$:\cite{nogu21a}
\begin{eqnarray}\label{eq:phi0}
\phi &=& \frac{1}{1+\exp(w_{\mathrm{b}})}, \\ \nonumber
w_{\mathrm{b}}&=& - \frac{\mu}{k_{\mathrm{B}}T}   + \frac{a_{\mathrm{p}}}{k_{\mathrm{B}}T}\Big[2(\kappa_{\mathrm{pi}}-\kappa_{\mathrm{d}})H^2  \\ && 
+ (\bar{\kappa}_{\mathrm{pi}}-\bar{\kappa}_{\mathrm{d}})K -2\kappa_{\mathrm{pi}}C_0H  + \frac{\kappa_{\mathrm{pi}}C_0^2}{2} \Big].
\end{eqnarray}
This relation reflects the detailed balance between protein binding and unbinding at a local membrane region:
$\eta_{\mathrm{ub}}/\eta_{\mathrm{b}}= \exp(w_{\mathrm{b}})$ for the kinetic equation
${\mathrm{d}}\phi/{\mathrm{d}}t = \eta_{\mathrm{b}}(1-\phi) - \eta_{\mathrm{ub}}\phi$.\cite{gout21,nogu22a}
For $b\ne 0$, $\phi$ can be solved iteratively by replacing $w_{\mathrm{b}}$ with $w_{\mathrm{b}} + 2b\phi a_{\mathrm{p}}/k_{\mathrm{B}}T$ in Eq.~(\ref{eq:phi0}).\cite{nogu21a}

For $\kappa_{\mathrm{pi}}>\kappa_{\mathrm{d}}$ ($\kappa_{\mathrm{pa}}>0$),
the protein density $\phi$ exhibits a peak at a finite curvature (referred to as the sensing curvature $H_{\mathrm{s}}$, see Fig.~\ref{fig:seniso}(a)).
The maximum value of $\phi$ increases from $0$ to $1$ with increasing $\mu$.
Notably, the protein binding differ between spherical and cylindrical membranes with the same mean curvature $H$,
when $\bar{\kappa}_{\mathrm{pi}} \ne \bar{\kappa}_{\mathrm{d}}$.
The proteins bind more to spherical membranes compared to cylindrical membranes
at ($\bar{\kappa}_{\mathrm{pi}} - \bar{\kappa}_{\mathrm{d}})/(\kappa_{\mathrm{pi}} - \kappa_{\mathrm{d}}) = -1$ (see Fig.~\ref{fig:seniso}(a)).
The sensing curvature $H_{\mathrm{s}}$ is obtained by solving ${\mathrm{d}}\phi/{\mathrm{d}}H =0$ using Eq.~(\ref{eq:phi0}) under the conditions $K=H^2$
for spherical membranes and $K=0$ for cylindrical membranes:
\begin{equation}\label{eq:hs}
H_{\mathrm{s}}= \frac{\kappa_{\mathrm{pi}}}{2\kappa_{\mathrm{dif}}}C_0, 
\end{equation}
where $\kappa_{\mathrm{dif}}=\kappa_{\mathrm{pi}}-\kappa_{\mathrm{d}}+(\bar{\kappa}_{\mathrm{pi}}-\bar{\kappa}_{\mathrm{d}})/2$
and $\kappa_{\mathrm{dif}}=\kappa_{\mathrm{pi}}-\kappa_{\mathrm{d}}$ for the spherical and cylindrical membranes, respectively.
As $\kappa_{\mathrm{pi}}$ increases from $\kappa_{\mathrm{d}}$ to $\infty$,
$H_{\mathrm{s}}$ decreases from $\infty$ to $C_0/(2+\bar{\kappa}_{\mathrm{pi}}/\kappa_{\mathrm{pi}})$ 
for spherical membranes and to $C_0/2$ for cylindrical membranes (see Fig.~\ref{fig:seniso}(b)).
It is important to note that $\phi(H)$ is mirror symmetric with respect to the sensing curvature for both cylindrical and spherical membranes (see Fig.~\ref{fig:seniso}(a)).

In contrast, for $\kappa_{\mathrm{pi}}<\kappa_{\mathrm{d}}$,
the bound membrane exhibits a lower bending rigidity compared to the bare membrane.
This scenario may arise when the bound proteins (or other molecules)
remodel the bound membrane. For example, a reduction in membrane thickness can lead to decreased bending rigidity.
Interestingly, the proteins hold a negative curvature sensing at $\kappa_{\mathrm{pi}}<\kappa_{\mathrm{d}}$, where
$\phi$ exhibits a minimum instead of a maximum (see the gray lines in Fig.~\ref{fig:seniso}(a)).\cite{nogu22a}
In other words, the fraction of bare membrane, $1-\phi$, reaches its maximum at the negative sensing curvature.
Owing to the lower rigidity, the bound membranes bend passively to reduce the bending energy of bare membrane regions,
sometimes even in the opposite direction to their spontaneous curvatures.
Consequently, these proteins cannot induce membrane bending to their spontaneous curvatures.
Therefore, a higher bending rigidity ($\kappa_{\mathrm{pi}}>\kappa_{\mathrm{d}}$) is required to bend membranes to a specific curvature.

For $\kappa_{\mathrm{pi}}=\kappa_{\mathrm{d}}$,
$\phi$ follows a monotonic sigmoid function of $H$ without any distinct peaks (see the green line in Fig.~\ref{fig:seniso}(a)).
In several previous studies,\cite{sorr12,shi15,tozz19,kris19}
the condition $\kappa_{\mathrm{pi}}=\kappa_{\mathrm{d}}$ was set as a simplified model,
and the following bending energy was used:
\begin{equation}\label{eq:Fcv2}
F_{\mathrm{cv1}} =  \int {\mathrm{d}}A \ \Big\{ \frac{\kappa_{\mathrm{d}}}{2}(2H- C_0\phi)^2  \Big\}.
\end{equation}
This formulation corresponds to the condition of $\kappa_{\mathrm{pi}}=\kappa_{\mathrm{d}}$,  $\bar{\kappa}_{\mathrm{pi}}=\bar{\kappa}_{\mathrm{d}}$, 
and $b=\kappa_{\mathrm{d}}C_0^2/2$. 
The quadratic term $(\kappa_{\mathrm{d}}C_0^2/2)\phi^2$ is often neglected.\cite{rama00,shlo09}
Since this quadratic term is independent of membrane curvature
and represents a pairwise inter-protein interaction,
its inclusion in the bending energy is not recommended.
Similarly, preaveraging both bending rigidity and spontaneous curvature 
as $F_{\mathrm{cv1}} =  \int {\mathrm{d}}A\  (\kappa_{\mathrm{d}}+\kappa_1\phi)(2H- C_0\phi)^2/2$
is not advisable, because
it implicitly accounts for pairwise and three-body inter-protein interactions ($(2\kappa_1C_0H+\kappa_{\mathrm{d}}C_0^2/2)\phi^2$ and $(\kappa_1C_0^2/2)\phi^3$, respectively).~\cite{nogu15b,nogu22a}
Although the previous studies~\cite{tozz19,has21,tsai21} have compared 
 the two models given by Eqs.~(\ref{eq:Fcv1}) and (\ref{eq:Fcv2}) as distinct approaches, they are, in fact,
 the subsets of Eq.~(\ref{eq:Fcv1}) for $\kappa_{\mathrm{pi}}\ne \kappa_{\mathrm{d}}$ and $\kappa_{\mathrm{pi}}=\kappa_{\mathrm{d}}$, respectively.

The chemical potential $\mu$ can be modulated by adjusting the bulk protein concentration $\rho$.
For a dilute solution, it is expressed as $\mu(\rho)=\mu(1) + k_{\mathrm{B}}T\ln(\rho)$.
In experiments, the ratio of surface protein densities at different curvatures
has  often been used, making the estimation of $\mu$ unnecessary.
For a large spherical vesicle with $R_{\mathrm{A}}C_0 \gg 1$,
the membrane can be approximated as flat ($H=K=0$),
and the protein density is given by $\phi_{\mathrm{flat}} = 1/\{1+\exp[(-\mu + a_{\mathrm {p}}\kappa_{\mathrm {pi}}C_0^2/2)/k_{\mathrm {B}}T]\}$ for $b=0$.
Hence, for the protein density $\phi_{\mathrm{cy}}$ in a cylindrical membrane with radius $R_{\mathrm{cy}}$,
Eq.~(\ref{eq:phi0}) can be rewritten as\cite{nogu23b}
\begin{equation}\label{eq:phiia}
\phi_{\mathrm{cy}} = \frac{1}{1+ \frac{1-\phi_{\mathrm{flat}}}{\phi_{\mathrm{flat}}}\exp\big[\frac{a_{\mathrm {p}}}{k_{\mathrm {B}}T}\big(\frac{\kappa_{\mathrm {pi}}-\kappa_{\mathrm {d}}}{{2R_{\mathrm{cy}}}^2}
 -\frac{\kappa_{\mathrm {pi}}C_0}{R_{\mathrm{cy}}} \big)\big] }.
\end{equation}
In the low-density limit ($\phi_{\mathrm{flat}}\ll 1$ and $\phi_{\mathrm{cy}}\ll 1$),
the density ratio is simplified to an exponential function as\cite{prev15,nogu21b}
\begin{eqnarray}\label{eq:phiis}
\frac{\phi_{\mathrm{cy}}}{\phi_{\mathrm{flat}}} &=& \exp\Big[ -\frac{a_{\mathrm {p}}}{k_{\mathrm {B}}T}\Big(\frac{\kappa_{\mathrm {pi}}-\kappa_{\mathrm {d}}}{{2R_{\mathrm{cy}}}^2}
 -\frac{\kappa_{\mathrm {pi}}C_0}{R_{\mathrm{cy}}} \Big)\Big] \\ \nonumber
 &=& \exp\Big[ -\frac{a_{\mathrm {p}}}{k_{\mathrm {B}}T}\Big(\frac{\kappa_{\mathrm {pa}}}{{2R_{\mathrm{cy}}}^2}
 -\frac{\kappa_{\mathrm {pa}}C_{\mathrm {0a}}}{R_{\mathrm{cy}}} \Big)\Big],
\end{eqnarray}
for the bending-energy formulations given in Eqs.~(\ref{eq:Fcv0}) and (\ref{eq:Fcv1}), respectively.
In this limit, the ratio $\phi_{\mathrm{cy}}/\phi_{\mathrm{flat}}$ is independent of $\phi_{\mathrm{flat}}$.

\subsection{Intrinsically Disordered Protein (IDP) Domains}\label{sec:idp}

Many curvature-inducing proteins contain IDP domains.
Stachowiak and coworkers have investigated the effects of varying the length of IDP domains in BAR and other proteins
and have reported that the disordered domains facilitate curvature sensing
and that the longer IDP chains promote the formation of small vesicles.\cite{stac12,busc15,zeno19,snea19}
A disordered domain behaves as a linear polymer chain in a good solvent,\cite{hofm12}
that is, its mean radius of gyration scales as $\langle R_{\mathrm{g}} \rangle \sim {n_{\mathrm{poly}}}^{\nu}$,
where  $\langle ... \rangle$ denotes the ensemble average,
$n_{\mathrm{poly}}$ represents the number of Kuhn segments, and $\nu=0.6$ is the scaling exponent for an excluded volume chain.\cite{doi86,stro97}
The Kuhn length of IDP chains is approximately $0.8$\,nm.\cite{hofm12}
The interactions between membrane-anchored polymer chains and membrane have been extensively studied through theory~\cite{lipo95a,hier96,bick01,mars03,bick06}, 
simulations~\cite{auth03,auth05,wern10,wu13}, and experiments~\cite{tsaf01,tsaf03,akiy03,niko07,koth11}.
The formation of spherical buds~\cite{tsaf01,tsaf03,niko07} and membrane tubes~\cite{tsaf03,akiy03,koth11} have been observed experimentally.
Polymer anchoring induces a positive spontaneous curvature of the membrane and increases the bending rigidity in a good solvent.

At low polymer densities (referred to as the ``mushroom regime''), the polymer chain exists in isolation on the membrane forming a mushroom-like distribution, where the inter-polymer interactions are negligible.
In this regime,
both the spontaneous curvature and  bending rigidity 
increase linearly with the grafting density $\phi_{\mathrm{poly}}$ of the polymer chains.
Analytically, the relations\cite{lipo95a,hier96}
\begin{eqnarray}
\kappa_{\mathrm{pi}} \Delta H_0 &=& \hspace{0.25cm} k_{\mathrm {h0}} k_{\mathrm{B}}T R_{\mathrm {end}} \phi_{\mathrm{poly}}, \label{eq:scvp} \\
\Delta \kappa &=& \hspace{0.25cm}  k_{\kappa}k_{\mathrm{B}}T {R_{\mathrm{end}}}^2 \phi_{\mathrm{poly}}, \label{eq:kappap} \\
\Delta \bar{\kappa} &=&   -\bar{k}_{\kappa}k_{\mathrm{B}}T {R_{\mathrm{end}}}^2 \phi_{\mathrm{poly}}, \label{eq:bkappap}
\end{eqnarray}
are predicted, where 
$\Delta H_0= \Delta C_0/2$, $\Delta \kappa$, and $\Delta \bar{\kappa}$ represent the
differences in the spontaneous curvatures, bending rigidities, and saddle-splay moduli between
the polymer-decorated and bare membranes, respectively, and
 $R_{\mathrm {end}}$ represents the mean end-to-end distance of the polymer chain.
These coefficients have been analytically derived
using Green's function for ideal chains \cite{lipo95a,hier96} and have also been
 estimated by Monte Carlo (MC) simulations of single anchored polymer chains~\cite{auth03}:
 $k_{\mathrm {h0}}=0.18$ and $0.17$; $k_{\kappa}=0.21$ and $0.2$; and  $\bar{k}_{\kappa}= 0.17$ and $0.15$;
for ideal and excluded-volume chains, respectively.

\begin{figure}[tbh]
\includegraphics[]{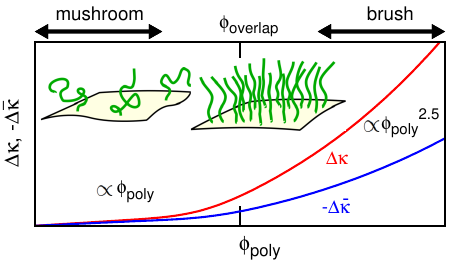}
\caption{
Schematic graph of the bending modulus modification by IDP chains.
The red and blue lines represent the differences in the bending rigidities ($\Delta\kappa$), and 
saddle-splay moduli ($\Delta\bar{\kappa}$) between the IDP-decorated and bare membranes, respectively.
At sufficiently lower and higher densities than the overlap density ($\phi_{\mathrm{overlap}}$),
the IDP chains exhibit mushroom and brush shapes, respectively, as drawn in the inset.
In the mushroom region, $\Delta\kappa$ and $\Delta\bar{\kappa}$ are expressed by Eqs.~(\ref{eq:kappap}) and (\ref{eq:bkappap}), respectively, with the coefficients for a good solvent.\cite{auth03}
In the brush region, $\Delta\kappa$ and $\Delta\bar{\kappa}$ are expressed by Eqs.~(\ref{eq:brush1}) and (\ref{eq:brush2}), respectively, in the small curvature limit.\cite{hier96}
}
\label{fig:brush}
\end{figure}

At a polymer density sufficiently higher than the overlap density (referred to as ``brush regime''), 
the polymer chains extend perpendicularly from the membrane surface, forming a brush-like structure.
In this regime, 
polymers grafting further enhance both the bending rigidity and spontaneous curvature of the membrane.
In the limit of small curvature,
the bending rigidity and saddle-splay modulus are given by\cite{hier96}
\begin{eqnarray}\label{eq:brush1}
\Delta\kappa &=& \frac{\nu+2}{12\nu^2} {n_{\mathrm{poly}}}^3{{\phi^*}_{\mathrm{poly}}}^{3/2\nu}k_{\mathrm{B}}T, \\
\Delta\bar{\kappa} &=& - \frac{1}{6\nu} {n_{\mathrm{poly}}}^3{{\phi^*}_{\mathrm{poly}}}^{3/2\nu}k_{\mathrm{B}}T,
\label{eq:brush2}
\end{eqnarray}
where ${\phi^*}_{\mathrm{poly}}$ is the polymer density normalized by the maximum coverage.
Consequently, brush polymers increases the membrane rigidity in proportion to ${\phi_{\mathrm{poly}}}^{2.5}$,
as show in Fig.~\ref{fig:brush}.

In addition,
polymer grafting reduces the line tension of membrane edges, thereby stabilizing the microdomains with a size of the polymer-chain length.\cite{wu13}
Furthermore, in a poor solvent environment, the polymer grafting can induce a negative spontaneous curvature, leading to the formation of a dimple-shaped membrane structure.\cite{wern10,evan03a}

\subsection{Theory of Anisotropic Proteins}\label{sec:anisens}

Here, we consider the binding of anisotropic proteins to membranes.
A prominent example of anisotropic proteins is the BAR superfamily proteins,
which features a banana-shaped binding domain known as the BAR domain. 
This binding domain is a dimer and holds twofold rotational symmetry.
The BAR domain binds to the membrane, inducing curvature along the domain axis and
generating cylindrical membrane tubes.\cite{mcma05,suet14,masu10,itoh06,mim12a,fros09,shi15,adam15,baro16}
The BAR domains have lengths from $13$ to $27$ nm.\cite{masu10}
N-BAR and F-BAR domains have a positive curvature along the domain axis
and I-BAR (Inverse-BAR) domains have a negative curvature.
Some BAR proteins also have extra binding domains, such as phox homology (PX) and pleckstrin homology (PH) domains,
and N-BAR and some extra domains have membrane insertion modules.\cite{itoh06,lemm08,chan19}
These domains and modules can modify the bending energy of bound regions, with maintaining the rotational symmetry.

Not all curvature-inducing proteins exhibit rotational symmetry.
For example, dynamin\cite{ferg12,anto16,pann18}, which has an asymmetric structure,
forms helical assemblies that constrict membrane neck, leading to membrane fission.
Similarly, melittin and amphipathic peptides\cite{sato06,rady17,guha19,miya22}
bind to membranes, and their circular assemblies result in membrane pore formation.
Recent coarse-grained simulation of a buckled membrane by
G{\'o}mez-Llobregat and coworkers demonstrated the curvature sensing of three amphipathic peptides.\cite{gome16}
They revealed that melittin and the amphipathic peptides LL-37 (PDB: 2k6O) exhibited asymmetric curvature sensing,
 meaning that their angular distribution relative to the buckled axis is not symmetric.

Several bending-energy models have been proposed to describe the behavior of
 anisotropic proteins.
For the crescent-shaped symmetric proteins, such as BAR proteins,
the bending energy can be expressed  as\cite{nogu22a,nogu16,tozz21}
\begin{equation} \label{eq:ubrod}
U_{\mathrm {p}} =   \frac{\kappa_{\mathrm{p}}a_{\mathrm{p}}}{2}(C_{\ell \mathrm{m}} - C_{\mathrm{p}})^2 + \frac{\kappa_{\mathrm{s}}a_{\mathrm{p}}}{2}(C_{\ell \mathrm{s}} - C_{\mathrm{s}})^2,
\end{equation}
where $\kappa_{\mathrm{p}}$ and $C_{\mathrm{p}}$ represent the bending rigidity and spontaneous curvature along the major protein axis, respectively,
while  $\kappa_{\mathrm{s}}$ and $C_{\mathrm{s}}$ denote those along the minor (side) axis.
The membrane curvatures along these major and minor axes are given by
\begin{eqnarray} \label{eq:cvl12}
C_{\ell \mathrm{m}} &=& C_1\cos^2(\theta_{\mathrm {pc}}) +C_2\sin^2(\theta_{\mathrm {pc}}) = H + D\cos(2\theta_{\mathrm {pc}}), \hspace{0.55cm} \\
C_{\ell \mathrm{s}} &=& C_1\sin^2(\theta_{\mathrm {pc}}) +C_2\cos^2(\theta_{\mathrm {pc}}) = H - D\cos(2\theta_{\mathrm {pc}}),
\end{eqnarray}
respectively,
where $D= (C_1-C_2)/2$ represents the deviatoric curvatures of the membrane ($D^2= H^2 - K$), and 
 $\theta_{\mathrm {pc}}$ represents the angle between the protein axis and the direction of one of the principal membrane curvatures
(typically, the azimuthal direction is selected for cylindrical membranes). 
The protein bends the membrane with $\kappa_{\mathrm{p}}$ and $C_{\mathrm{p}}$ along the major axis
and with $\kappa_{\mathrm{s}}$ and $C_{\mathrm{s}}$ along the minor axis.
If the side regions of the linear-shaped proteins bind strongly to the membrane,
they exhibit a negative side curvature ($C_{\mathrm{s}}<0$).\cite{simu13,olin16}
Conversely, the excluded-volume repulsion between adjacent proteins can generate a positive side curvature.\cite{nogu17}

\begin{figure}[]
\includegraphics[]{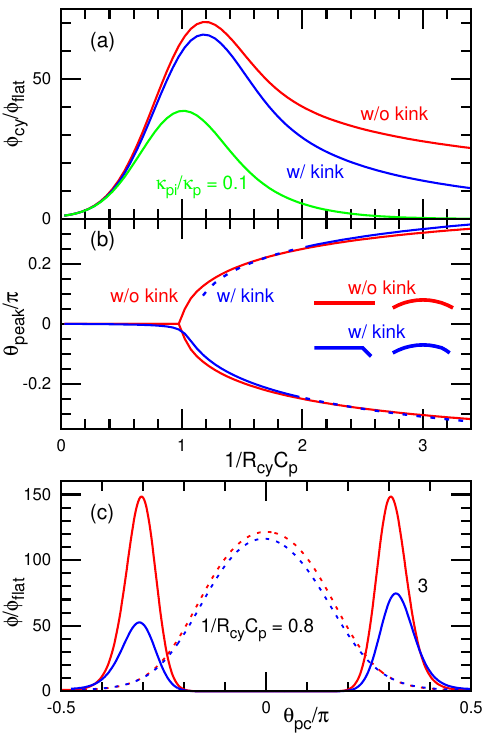}
\caption{
Binding of anisotropic proteins in the low-density limit with $C_{\mathrm{p}}^2a_{\mathrm{p}} = 0.2$, $\kappa_{\mathrm{p}}=50k_{\mathrm{B}}T$, and $\kappa_{\mathrm{s}}=0$.\cite{nogu24}
The red lines represent the data of a twofold  rotationally symmetric protein (crescent-rod shape without kinks).
The blue lines represent the data of an asymmetric  protein, where
the rod-shaped protein bends at a kink with an angle of $\pi/4$, 
positioned at 20\% of the protein length from the end.
The axis of the asymmetric  protein is set to be $\theta_{\mathrm{peak}}=0$ at $1/R_{\mathrm{cy}}\ll 1$.
The green line represents the data of the twofold rotationally symmetric protein 
with an isotropic segment of $\kappa_{\mathrm{pi}}/\kappa_{\mathrm{p}}=0.1$.
(a) Binding density $\phi_{\mathrm{cy}}$ on a cylindrical membrane with respect to 
the density $\phi_{\mathrm{flat}}$ on a flat membrane.
(b) Peak position of the angle $\theta_{\mathrm{pc}}$.
The solid and dashed lines represent the first and second peaks, respectively.
The inset shows the schematics of the top and side views of proteins.
(c) Distribution of the angle $\theta_{\mathrm{pc}}$.
The solid and dashed lines represent the data for $1/R_{\mathrm{cy}}C_{\mathrm{p}}=3$ and $0.8$, respectively.
}
\label{fig:sen1}
\end{figure}

A protein can comprise binding domains with distinct bending axes (where $C_{\ell j}$ denotes the membrane curvature along the axis of the $j$-th domain)
and isotropic bending regions (IDP domains etc.).
Consequently, the bending energy of a single protein is generally expressed as\cite{nogu24}
\begin{eqnarray} \label{eq:cvg0}
U_{\mathrm {p}} 
=&& F_{\mathrm{pi}} + \sum_j^{N_{\mathrm{ax}}} \frac{\kappa_{\mathrm{p}j}a_{\mathrm{p}}}{2}(C_{\ell j} - C_{\mathrm{p}j})^2 \\ \label{eq:cvg1}
=&& k_1 H^2 + k_2 H + k_3 K + k_4 D\cos(2\theta_{\mathrm {pc}})  \nonumber \\ \nonumber 
&+& k_5 HD\cos(2\theta_{\mathrm {pc}}) + k_6 D^2\cos(4\theta_{\mathrm {pc}}) \\ \nonumber 
&+& k_7 D\sin(2\theta_{\mathrm {pc}}) + k_8 HD\sin(2\theta_{\mathrm {pc}}) \\
&+& k_9 D^2\sin(4\theta_{\mathrm {pc}}) + U_0,
\end{eqnarray}
in the second-order expansion of membrane curvature.
The constant term $U_0$ can be neglected by incorporating it into the chemical potential, such that $\mu'= \mu + U_0$.
Isotropic proteins are characterized by the first three terms with the coefficients $k_1=2\kappa_{\mathrm {pa}}a_{\mathrm {p}}$, $k_2=-2\kappa_{\mathrm {pa}}a_{\mathrm {p}} C_{\mathrm {0a}}$, and $k_3 = \bar{\kappa}_{\mathrm {pa}}a_{\mathrm {p}}$ for $F_{\mathrm{pi}}$ (compare Eqs.~(\ref{eq:Fcv1}) and (\ref{eq:cvg1})).
Proteins with twofold rotational or mirror symmetry can have the first six terms ($k_1$--$k_6$),
while asymmetric proteins may exhibit all nine terms.
The protein major axis can be chosen to be $k_7=0$ in order to reduce the number of coefficients.
To express asymmetry, $k_8\ne 0$ or $k_9\ne 0$ is needed at this axis.
The protein model in Eq.~(\ref{eq:ubrod}) is considered with $N_{\mathrm{ax}}=2$ and $F_{\mathrm{pi}}=0$, 
assuming orthogonal axes where $C_{\ell 1} = C_{\ell \mathrm{m}}$ and $C_{\ell 2} = C_{\ell \mathrm{s}}$, 
and the coefficients are mapped accordingly as $k_1= 3(\kappa_{\mathrm {p}}+\kappa_{\mathrm {s}})a_{\mathrm {p}}/4$,
$k_2= - (\kappa_{\mathrm {p}}C_{\mathrm{p}}+\kappa_{\mathrm {s}}C_{\mathrm{s}})a_{\mathrm {p}}$,
$k_3= -(\kappa_{\mathrm {p}}+\kappa_{\mathrm {s}})a_{\mathrm {p}}/4$,
$k_4= -(\kappa_{\mathrm {p}}C_{\mathrm{p}}-\kappa_{\mathrm {s}}C_{\mathrm{s}})a_{\mathrm {p}}$,
$k_5= (\kappa_{\mathrm {p}}-\kappa_{\mathrm {s}})a_{\mathrm {p}}$, and
$k_6= (\kappa_{\mathrm {p}}+\kappa_{\mathrm {s}})a_{\mathrm {p}}/4$.\cite{nogu24}
Akabori and Santangelo\cite{akab11} have added 
$U_{\mathrm {asy}} = k_{\mathrm {asy}}[ D\sin(2\theta_{\mathrm {pc}}) - C_{\mathrm {asy}}]^2$
to Eq.~(\ref{eq:ubrod}) in order to include an asymmetric bending effect.
Their formulation corresponds to Eq.~(\ref{eq:cvg1}) with $k_7= -2k_{\mathrm {asy}}C_{\mathrm {asy}}$ and $k_8=k_9=0$, modifying $k_1$, $k_3$, and $k_6$.
{Kralj-Igli\v{c}} and coworkers have considered  
the protein energy with a symmetric shape,\cite{kral99,igli07}
$U_{\mathrm {p}} = k_{\mathrm {a}}(H - H_0)^2/2  + (k_{\mathrm {a}}+k_{\mathrm {b}})[D^2 - 2DD_0\cos(2\theta_{\mathrm {pc}})  + {D_0}^2]/4$.
The second term assumes an energy proportional to a rotational average of $({\mathrm{d}}(C_\ell-C_m)/{\mathrm{d}}\theta)^2$,
where $C_\ell$ is the normal membrane curvature at the angle $\theta$, and $C_m= C_{m0}+ C_{m1}\cos(2\theta)$ is the angle-dependent spontaneous curvature.
In this formulation, $k_1= 3k_{\mathrm {a}}/4+k_{\mathrm {b}}/4$, $k_2= -k_{\mathrm {a}}H_0$, $k_3= -(k_{\mathrm {a}}+k_{\mathrm {b}})/4$,
$k_4= -(k_{\mathrm {a}}+k_{\mathrm {b}})D_0/2$, and $k_5=k_6=0$. They have also used the first term of Eq.~(\ref{eq:ubrod}) for rod-like proteins
(i.e., $\kappa_{\mathrm{s}}=0$).\cite{igli07}
Fournier combined an anisotropic bending energy with the tilt energy of lipids for transmembrane proteins.\cite{four96}

\subsubsection{Isolated Proteins}\label{sec:anilow}

First, we consider protein binding in the low-density limit,
in which bound proteins are isolated on the membrane and inter-protein interactions are negligible.
Hence, the density $\phi$ of bound proteins is given by $\phi= (1/2\pi)\int_{-\pi}^{\pi} \exp[(\mu-U_{\mathrm {p}})/k_{\mathrm {B}}T]\ {\mathrm d}\theta_{\mathrm {pc}}$.
The binding ratio of proteins to a cylindrical membrane tube relative to a flat membrane
is expressed as\cite{nogu24}
\begin{equation}\label{eq:phia}
\frac{\phi_{\mathrm {cy}}}{\phi_{\mathrm {flat}}} = \frac{\exp\big(\frac{U_{\mathrm {p}}^{\mathrm {flat}}}{k_{\mathrm {B}}T}\big)}{2\pi}\int_{-\pi}^{\pi} \exp\Big(-\frac{U_{\mathrm {p}}^{\mathrm {cy}}}{k_{\mathrm {B}}T}\Big)\ {\mathrm d}\theta_{\mathrm {pc}},
\end{equation}
where $U_{\mathrm {p}}^{\mathrm {flat}}$ is the bending energy for the flat membrane, and 
$U_{\mathrm {p}}^{\mathrm {cy}}$ is that for the cylindrical membrane.
This ratio $\phi_{\mathrm {cy}}/\phi_{\mathrm {flat}}$ is independent of $\mu$ in the low-density limit, as in the isotropic proteins. 

Anisotropic proteins can adjust their lateral orientation to reduce their bending energy.
Let us consider a crescent symmetric protein (Eq.~(\ref{eq:ubrod}) with $\kappa_{\mathrm {s}}=0$) and its variants as simple anisotropic protein models.
This crescent protein has the lowest bending energy at $\theta_{\mathrm{pc}}=0$ (the protein orients in the azimuthal direction)
in a wide cylinder ($1/R_{\mathrm {cy}}C_{\mathrm {p}} \le 1$), 
whereas tilt proteins have the lowest at $\theta_{\mathrm{pc}}=\pm \arccos(\sqrt{R_{\mathrm{cy}}C_{\mathrm{p}}})$ in a narrow cylinder ($1/R_{\mathrm {cy}}C_{\mathrm {p}} > 1$).
Hence, the protein density exhibits peaks at these preferred orientations (see the red lines in Fig.~\ref{fig:sen1}(b) and (c)).
The average density $\phi_{\mathrm {cy}}$  also exhibits a peak at a membrane curvature slightly higher than $1/R_{\mathrm {cy}}C_{\mathrm {p}} = 1$ (see Fig.~\ref{fig:sen1}(a)).
Unlike isotropic proteins, $\phi_{\mathrm {cy}}(1/R_{\mathrm {cy}})$ is not mirror symmetric and decreases gradually at larger curvatures,
owing to the angular adjustment of proteins.
When an isotropic bending energy component, $F_{\mathrm{pi}}$, is added with a relative strength of 10\% ($\kappa_{\mathrm{pi}}/\kappa_{\mathrm{p}}=0.1$ and $C_{\mathrm{0a}}=0$),
the density profile of $\phi_{\mathrm {cy}}$ approaches a mirror symmetric shape (see the green line in Fig.~\ref{fig:sen1}(a)).
Some amphipathic peptides have a kink structure, which allows significant bending.
To mimic this behavior, a kink is introduced at  20\% of the protein length from the protein end; at the kink, the protein bends laterally at an angle of  $\pi/4$.
Owing to the resulting asymmetry, the angular distribution becomes skewed, with the highest peak appearing at $\theta_{\mathrm {pc}}<0$ and $\theta_{\mathrm {pc}}>0$ for the curvature ranges
$1< 1/R_{\mathrm {cy}}C_{\mathrm {p}} < 2$ and $1/R_{\mathrm {cy}}C_{\mathrm {p}} > 2$, respectively (see the blue lines in Fig.~\ref{fig:sen1}).\cite{nogu24}
A similar asymmetric angular distribution was reported in molecular simulation.\cite{gome16}
The above discussion focuses on the binding of rigid proteins;
however, the deformation of the binding domains can modify the protein density as demonstrated in Ref.~\citenum{nogu24}.

\subsubsection{High Protein Density}\label{sec:aniden}

Next, we describe a mean-field theory\cite{tozz21,nogu22}
that accounts for orientation-dependent excluded area,
in which Nascimentos' theory\cite{nasc17} for three-dimensional (3D) liquid-crystals is applied to the 2D membrane.
Bound proteins are assumed to adopt an elliptical shape laterally on the membrane
and can be aligned based on their inter-protein interactions and their preferred bending direction.
The degree of orientational order $S$ is given by
$S = 2 \langle s_{\mathrm{p}}(\theta_{\mathrm {ps}}) \rangle$, where
$s_{\mathrm {p}}(\theta_{\mathrm {ps}}) = \cos^2(\theta_{\mathrm{ps}}) - 1/2$
and  $\theta_{\mathrm{ps}}$ denotes the angles between the major protein axis and nematic orientation {\bf S}.
The protein area is defined as $a_{\mathrm{p}} = \pi \ell_1\ell_2/4$,
where $\ell_1$ are $\ell_2$ represent the lengths of the major and minor protein axes, respectively.

The free energy $F_{\mathrm{p}}$ of bound proteins is expressed as\cite{tozz21}
\begin{eqnarray}\label{eq:fpa1}
F_{\mathrm{p}} &=&  \int f_{\mathrm{p}}\ {\mathrm{d}}A, \\ \label{eq:fpa2}
f_{\mathrm{p}} &=&  \frac{\phi k_{\mathrm{B}}T}{a_{\mathrm{p}}}\Big[\ln(\phi) + \frac{S \Psi}{2} - \ln\Big(\int_{-\pi/2}^{\pi/2} w(\theta_{\mathrm{ps}})\ {\mathrm{d}}\theta_{\mathrm{ps}}\Big)\Big],\hspace{0.6cm}
\end{eqnarray}
where
\begin{eqnarray}
w(\theta_{\mathrm{ps}})  &=&  g\exp\big[\Psi s_{\mathrm{p}}(\theta_{\mathrm{ps}}) + \bar{\Psi}\sin(\theta_{\mathrm{ps}})\cos(\theta_{\mathrm{ps}}) \nonumber \\ && - U_{\mathrm{p}}/k_{\mathrm{B}}T \big]\Theta(g), \\
g   &=& 1-\phi [b_0-b_2S s_{\mathrm{p}}(\theta_{\mathrm{ps}})]. \label{eq:fpa3}
\end{eqnarray}
$\Psi$ and $\bar{\Psi}$ represent the symmetric  and asymmetric components of the nematic tensor, respectively. 
The factor $g$ accounts for the weight of the orientation-dependent excluded volume interaction,
and  $\Theta(x)$ denotes the unit step function.
When two proteins are aligned parallel to each other,
the excluded area $A_{\mathrm{exc}}$ between them is smaller compared to when they are oriented perpendicularly.
This difference increases with increasing aspect ratio $d_{\mathrm{el}}=\ell_1/\ell_2$.
The area  $A_{\mathrm{exc}}$ can be approximated as $A_{\mathrm {exc}}= [b_0 - b_2(\cos^2(\theta_{\mathrm{pp}})-1/2)]a_{\mathrm{p}}/\lambda$,
where $\theta_{\mathrm{pp}}$ is the angle between the major axes of two proteins, and
$\lambda$ represents the packing ratio.
The maximum density is given by  $\phi_{\mathrm{max}}=1/\lambda(b_0-b_2/2)$.\cite{nogu22}

For a flat membrane, proteins exhibit an isotropic orientation at low densities
and a first-order transition to a nematic order  at high densities
 owing to the orientation-dependent excluded volume interactions.\cite{tozz21}
In this review, we consider the anisotropic bending energy described by Eq.~(\ref{eq:ubrod}) with $\kappa_{\mathrm{s}}=0$ for $U_{\mathrm{p}}$.
As the curvature $1/R_{\mathrm{cy}}$ of a membrane tube increases,
proteins tend to align in the azimuthal direction even in the dilute limit (see Fig.~\ref{fig:sen1}(c)),
and the transition to the nematic state becomes continuous.

For narrow tubes with $1/R_{\mathrm{cy}}> C_{\mathrm{p}}$,
the preferred protein orientation tilts away from the azimuthal direction.
At low $\phi$, proteins tilted in both the left and right directions coexist equally (Fig.~\ref{fig:sen1}).
However, at high protein densities, only one type of tilt direction dominates due to 
orientation-dependent excluded volume interactions.
Thus, second-order and first-order transitions occur between these two states at medium and large curvatures, respectively.\cite{nogu22}

This theory well reproduces the simulation results for crescent protein rods on a membrane tube,
when the proteins are isotropically distributed.\cite{nogu22}
However, the discrepancies arise when the proteins form a significant amount of clusters,
since the current theory does not account for inter-protein attraction and assumes a homogeneous protein distribution.\cite{nogu22}

\begin{figure}[tbh]
\includegraphics[]{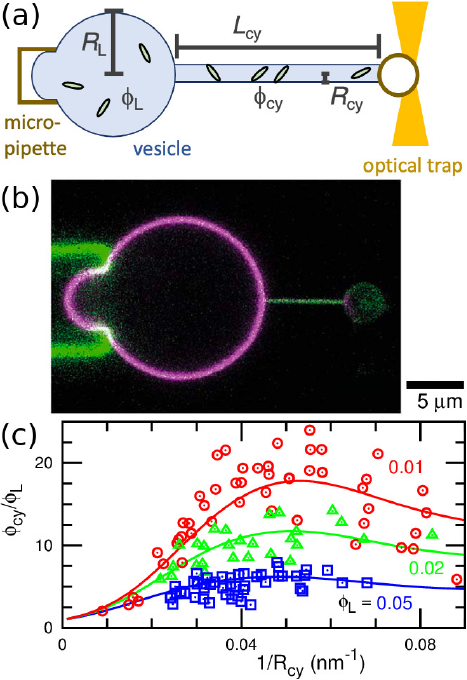}
\caption{
Binding of I-BAR domain of IRSp53 to tethered vesicle.
(a) Schematic of the experimental setup.
A cylindrical membrane tube (tether) is extended by an optical trap and micropipette.
(b) Confocal image of a vesicle with a tube of $R_{\mathrm{cy}}=25$\,nm.
Green and magenta indicate the fluorescence for I-BARs and lipids, respectively.
(c) Protein density $\phi_{\mathrm{cy}}$ in the tube normalized by that of the large spherical region $\phi_{\mathrm{L}}$.
Circles, triangles, and squares indicate
the experimental data of $\phi_{\mathrm{cy}}/\phi_{\mathrm{L}}$ for $\phi_{\mathrm{L}}=0.01$, $0.02$, and $0.05$, respectively.
The solid lines are obtained using fitting by the anisotropic protein model
with $\kappa_{\mathrm{p}}/k_{\mathrm{B}}T=82$ and $C_{\mathrm{p}}= -0.047$\,nm$^{-1}$.   
The experimental data in (b) and (c) are reproduced from Ref.~\citenum{prev15}. Licensed under CC BY (2015).
The plot in (c) is reproduced from Ref.~\citenum{nogu23b}  with permission from the Royal Society of Chemistry (2023).
}
\label{fig:ibar}
\end{figure}

\begin{figure}[tbh]
\includegraphics[]{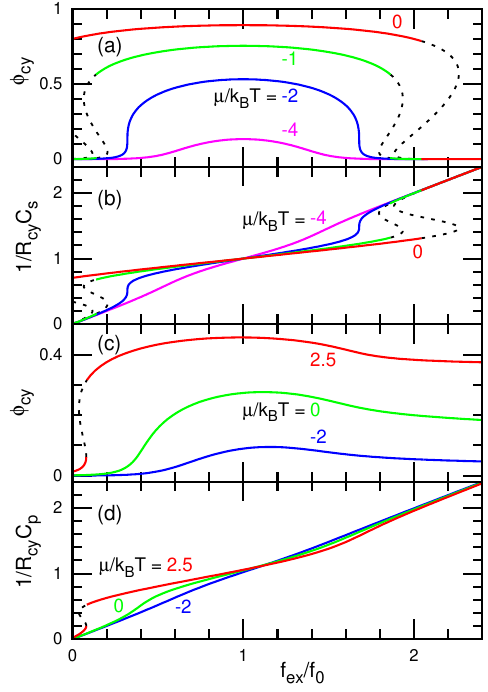}
\caption{
Protein binding to a membrane tube pulled by an external force $f_{\mathrm{ex}}$.
Protein density $\phi$ and tube curvature $1/R_{\mathrm{cy}}$ are shown in (a),(c) and in (b),(d), respectively.
(a),(b) Isotropic proteins for $\mu/k_{\mathrm{B}}T=-4$, $-2$, $-1$, and $0$ at $\kappa_{\mathrm{p}}/\kappa_{\mathrm{d}}= 4$
and $a_{\mathrm{p}}C_0^2 = 0.16$. 
(c),(d) Crescent elliptic proteins for $\mu/k_{\mathrm{B}}T=-2$, $0$, and $2.5$
  at $\kappa_{\mathrm{p}}/k_{\mathrm{B}}T=60$, $\kappa_{\mathrm{s}}=0$, $d_{\mathrm{el}}=3$, and  $a_{\mathrm{p}}{C_{\mathrm{p}}}^2 = 0.26$. 
The solid lines represent equilibrium states.
The black dashed lines represent metastable and free-energy barrier states (van der Waals loops).
The isotropic proteins exhibit a first-order transition twice at large $\mu$.\cite{nogu21b}
 In contrast, the anisotropic proteins exhibit it only once at a small curvature.\cite{nogu22}
}
\label{fig:fex}
\end{figure}

\subsection{Binding to Tethered Vesicle}\label{sec:tether}

A vesicle held by a micropipette forms a narrow membrane tube (tether) under a pulling force applied by optical tweezers,
 as illustrated in Fig.~\ref{fig:ibar}(a).\cite{dimo14,bo89,evan96,cuve05}
The tube radius can be controlled by adjusting the force strength.
Tethered vesicles have been widely employed to study the curvature sensing of membrane proteins, including
BAR proteins,\cite{baum11,has21,sorr12,prev15,tsai21} 
ion channels,\cite{aimo14,yang22} GPCRs,\cite{rosh17} dynamin,\cite{roux10}  annexins,\cite{more19} and Ras proteins.\cite{lars20} 

Protein density in the membrane can be quantified using fluorescence intensity measurement, as shown in Fig.~\ref{fig:ibar}(b).
For I-BAR domains, the density ratio $\phi_{\mathrm{cy}}/\phi_{\mathrm{L}}$ between the membrane tube to large spherical regions reaches a peak at a tube curvature of approximately $0.05$\,nm$^{-1}$ and gradually decreases at larger curvature (see Fig.~\ref{fig:ibar}(c)).\cite{prev15}
This curvature dependence can be reproduced by the theory for elliptic proteins (Eqs.~(\ref{eq:fpa1})--(\ref{eq:fpa3})) with
 $\kappa_{\mathrm{p}}/k_{\mathrm{B}}T=82$, $C_{\mathrm{p}}= -0.047$\,nm$^{-1}$, and $\kappa_{\mathrm{s}}=0$.\cite{nogu23b} 
Note that the theory for isotropic proteins (Eq.~(\ref{eq:phiia}) or (\ref{eq:phiis})) can reproduce each curve using different $\kappa_{\mathrm{pi}}$ and $C_0$\cite{prev15} but cannot simultaneously fit all three experimental curves.\cite{nogu23b}
This finding strongly supports the anisotropic nature of the curvature sensing in I-BAR domains.
Therefore, the tethered vesicle serves as a valuable tool not only for investigating curvature sensing but also for estimating the bending properties of various membrane proteins. 
However, the dependence on the saddle-splay modulus ($k_3$ in Eq.~(\ref{eq:cvg1})) cannot be directly measured using the tethered vesicle, 
since $K=0$ in the membrane tube.
Instead, $k_3$ can be estimated by comparing curvature sensing data from the membrane tubes and spherical vesicles with the same mean curvature (see Fig.~\ref{fig:seniso}).
Curvature sensing has been observed through protein binding to spherical vesicles with various sizes,\cite{lars20,hatz09,zeno19}
and the comparisons with the data in membrane tube were also reported in Ref.~\citenum{lars20} at small membrane curvatures.
For the estimation of the protein properties, the sensing data at large curvatures are particularly significant,
since the anisotropic characteristics become more pronounced in this regime (see Fig.~\ref{fig:sen1}(a)).

The force generated by the bending energy, while maintaining a fixed volume and surface area,
is balanced with the external force $f_{\mathrm {ex}}$ at equilibrium.
Under typical experimental conditions of the tethered vesicle,
the membrane tube is extremely narrow,
making the volume of the cylindrical tube 
negligible, as $R_{\mathrm {cy}}^2L_{\mathrm{cy}}/{R_{\mathrm{A}}}^3 \ll 1$.\cite{smit04,nogu21b} 
In this limit condition,
the vesicle shape is obtained from $\partial F_{\mathrm{cv1}}/\partial L_{\mathrm{cy}}=f_{\mathrm {ex}}|_{A_{\mathrm{cy}}}$
of the cylindrical tube with $A_{\mathrm{cy}}= 2\pi R_{\mathrm{cy}} L_{\mathrm{cy}}$.

For the binding of isotropic proteins, it is expressed as\cite{nogu21b} 
\begin{equation}  \label{eq:fext1}
f_{\mathrm {ex}} \approx
 \frac{2\pi[ (\kappa_{\mathrm {p}}-\kappa_{\mathrm {d}})\phi_{\mathrm{cy}}+\kappa_{\mathrm{d}}]}{R_{\mathrm {cy}}}
- 2\pi\kappa_{\mathrm{p}}C_0\phi_{\mathrm {cy}},
\end{equation}
where $\phi_{\mathrm {cy}}$ is given by Eq.~(\ref{eq:phi0}).
For the bare membrane ($\phi_{\mathrm {cy}}=0$), 
a linear relation is obtained between the force and the tube curvature as
$f_{\mathrm {ex}}=2\pi \kappa_{\mathrm{d}}/R_{\mathrm{cy}}$, which
is widely used to estimate the bending rigidity of the bare membrane.\cite{kara23,dimo14,mars06,rawi00,bo89}
The protein density $\phi_{\mathrm{cy}}$ and the tube curvature $1/R_{\mathrm{cy}}$ exhibit mirror and point symmetry
 with respect to $f_{\mathrm {ex}}/f_0=1$, as shown in Fig.~\ref{fig:fex}(a) and (b), respectively, 
where $f_0=2\pi \kappa_{\mathrm{d}}C_{\mathrm{s}}$ represents the force at the sensing curvature $C_{\mathrm{s}}=2H_{\mathrm{s}} = \kappa_{\mathrm{pi}}C_0/(\kappa_{\mathrm{pi}}-\kappa_{\mathrm{d}})$.
At high $\mu$, a first-order transition occurs twice symmetrically at both weak and strong forces $f_{\mathrm {ex}}$ (see the red and green curves in Fig.~\ref{fig:fex}(a) and (b)).
At the transition point, narrow and wide tubes with different protein densities can coexist. 

For the anisotropic proteins,
the membrane curvature is obtained from the force balance as
$f_{\mathrm {ex}}/2\pi = \partial f_{\mathrm{p}}/\partial(1/R_{\mathrm{cy}})|_{\phi_{\mathrm{cy}}} + \kappa_{\mathrm{d}}/R_{\mathrm{cy}}$,
where $f_{\mathrm{p}}$ is given by Eq.~(\ref{eq:fpa2}).\cite{nogu22,nogu23b}
The $f_{\mathrm{ex}}$ dependence curves of  $\phi_{\mathrm {cy}}$ and $1/R_{\mathrm{cy}}$ are not symmetric, unlike for isotropic proteins
(compare Fig.~\ref{fig:fex}(c) and (d) with Fig.~\ref{fig:fex}(a) and (b)).
The density and curvature exhibit a weaker dependence on $f_{\mathrm{ex}}$ at $f_{\mathrm{ex}}>f_0$ owing to the protein tilting in narrow tubes,
where $f_0= 2\pi \kappa_{\mathrm {d}}C_{\mathrm{p}}$.
Consequently, at high $\mu$, the first-order transition occurs only once in wide tubes.
This transition has been experimentally observed, showing the coexistence of high and low I-BAR density regions within the same membrane tube  in Ref.~\citenum{prev15}.
The sensing curvature of anisotropic proteins is influenced not only by $C_{\mathrm{p}}$ but also by the protein density,
as shown in Fig.~\ref{fig:fex}(c) and (d).\cite{nogu22}

\begin{figure}[t]
\includegraphics[]{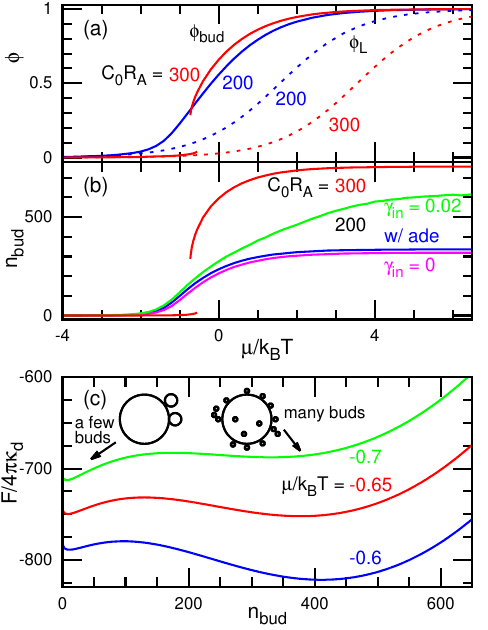}
\caption{
Budding of a vesicle induced by the binding of proteins with a spontaneous curvature $C_0$ at $V^*=0.95$, 
$\kappa_{\mathrm{pi}}/\kappa_{\mathrm{d}}=4$, and $\bar{\kappa}_{\mathrm{pi}}/\bar{\kappa}_{\mathrm{d}}=1$.\cite{nogu21a}
(a) Protein density on the vesicle surface in the absence of the ADE energy.
The solid and dashed lines represent the densities 
in the buds $\phi_{\mathrm{bud}}$ and spherical vesicle region $\phi_{\mathrm{L}}$, respectively.
The blue and red lines indicate the continuous and discontinuous transitions at $C_0R_{\mathrm{A}}=200$ and $300$, respectively.
(b) Number $n_{\mathrm{bud}}$ of buds. 
The blue and red lines represent the data at $C_0R_{\mathrm{A}}=200$ and $300$, respectively, in the absence of the ADE energy (corresponding to the data shown in (a)).
The green and magenta lines  represent the data with the ADE energy
in the presence and absence of protein insertion (the insertion area ratio of the protein $\gamma_{\mathrm{in}}=0.02$ and $0$), respectively, at $R_{\mathrm{A}}/h=5000$.
(c) Free energy profiles at $\mu/k_{\mathrm{B}}T=-0.7$, $-0.65$, and $-0.6$ (from top to bottom)
at $C_0R_{\mathrm{A}}=300$ (corresponding to the red lines in (a)).
Two minima for a few and many buds appear around the transition point.
}
\label{fig:bud}
\end{figure}

\begin{figure}[t]
\includegraphics[]{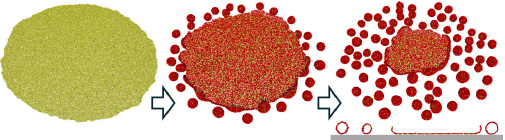}
\caption{
Sequential snapshots of membrane detachment from a substrate induced by the binding of isotropic proteins
at $C_0\sigma= 0.2$, $\kappa_{\mathrm{pi}}/k_{\mathrm{B}}T=34$, $\kappa_{\mathrm{d}}/k_{\mathrm{B}}T=16$, and $\mu/k_{\mathrm{B}}T =5$.\cite{nogu19c}
Detached membranes form small vesicles.
A sliced snapshot from the side view is shown for the right bird's-eye view snapshot.
The red and yellow spheres represent the membrane particles with and without the protein binding, respectively.
In the side view, the light gray rectangle represents the substrate.
}
\label{fig:disk}
\end{figure}

\section{Curvature generation}\label{sec:gen}

\subsection{Isotropic Proteins}\label{sec:isogen}

Curvature-inducing proteins alter the local membrane curvature, bringing it closer to their preferred curvatures.
In the absence of constraints,
the curvature $H_{\mathrm{g}}$ generated by isotropic proteins 
is determined by minimizing the free-energy, given by the condition ${\mathrm{d}}F_{\mathrm{cv1}}/{\mathrm{d}}H =0$ 
using Eq.~(\ref{eq:Fcv0}):\cite{nogu21a} 
\begin{equation}\label{eq:hg}
H_{\mathrm{g}}= \frac{\kappa_{\mathrm{pi}}\phi}{2(\kappa_{\mathrm{dif}}\phi + \kappa_{\mathrm{d}})}C_0, 
\end{equation}
where $\kappa_{\mathrm{dif}}$ represents the bending-rigidity difference as used in Eq.~(\ref{eq:hs}).
Since the proteins bend the underlying membrane together,
$H_{\mathrm{g}}$ is smaller than the sensing curvature $H_{\mathrm{s}}$ and
 depends on the membrane rigidity $\kappa_{\mathrm{d}}$, unlike the sensing curvature (see Fig.~\ref{fig:seniso}(b)). 
Moreover, $H_{\mathrm{g}}$ differs between spherical and cylindrical membranes at $\bar{\kappa}_{\mathrm{pi}} \ne \bar{\kappa}_{\mathrm{d}}$ 
(see Fig.~\ref{fig:seniso}(b)).
In the presence of constraints, the membrane may bend to a lesser extent than this generation curvature,
since the constraints can suppress the membrane deformation.

\subsubsection{Budding and Vesicle Formation}\label{sec:bud}

As an example of constraints, we consider the budding of a vesicle induced by protein binding.
Here, the volume and surface area of the vesicle is fixed (constrained) so that local membrane deformation 
maintains these constraints by entailing deformation in other membrane regions. 

In living cells, spherical buds typically form during vesicle formation.
In clathrins-mediated endocytosis, 
clathrins assemble on the membrane, forming spherical buds with diameters ranging from $20$ to $200$-nm.\cite{mcma11,kaks18,avin15,sale15}
Similarly, in the membrane trafficking between the endoplasmic reticulum and the Golgi apparatus,
COPI and COPII coated vesicles with diameters ranging from $60$ to $100$-nm are generated through budding under typical conditions.\cite{bran13,fain13,beth18}
These proteins can be considered as laterally isotropic, and
their budding processes have been theoretically analyzed using a spherical-cap geometry\cite{frey24,lipo92,sens03} 
and more detailed geometry.\cite{fore14}

The budding of a vesicle can be understood using the mean-field theory with simplified geometries.\cite{nogu21a}
A budded vesicle is modeled as small spheres connected to a large spherical membrane, 
as depicted in the inset of Fig.~\ref{fig:bud}(c).
Assuming that all buds have the same radius $R_{\mathrm{bud}}$,
the free energy minimum can be easily solved using Eq.~(\ref{eq:F0}) for one degree of freedom,
since the other two lengths can be determined by the area and volume constraints.
A prolate vesicle can be modeled by a cylinder shape capped with two hemispheres.
As the chemical potential $\mu$ increases, the protein density $\phi_{\mathrm {bud}}$ in the buds increases greater
than $\phi_{\mathrm{L}}$ in the large spherical region,
leading to the formation of a greater number of buds with a smaller radius (see Fig.~\ref{fig:bud}).
At a small spontaneous curvature ($C_0R_{\mathrm{A}}=200$),
The number of buds increases continuously, whereas, at a large spontaneous curvature ($C_0R_{\mathrm{A}}=300$),
a first-order transition occurs from a few buds with a large radius
to many buds with a small radius, as shown in Fig.~\ref{fig:bud}.
Thus, many buds can suddenly form after a long incubation period at slightly higher than the transition point.

This simplified geometrical framework can be easily applied to other shape transformations 
and is useful for investigating the effects of additional interactions.
For instance, the ADE energy is incorporated into the budding process (see Fig.~\ref{fig:bud}(b)).
Initially, the ADE energy is considered to be relaxed in the prolate vesicle ($\Delta A =\Delta A_0$ in the prolate).
When the bound proteins do not change $\Delta A_0$,
the ADE energy only slightly reduces the budding (see the magenta curve in Fig.~\ref{fig:bud}(b)).\cite{nogu21a}
However, the insertion of hydrophobic segments into the membrane can modify $\Delta A_0$.
When the segments insert only the outer leaflet with the ratio $\gamma_{\mathrm{in}}$ of the inserted area
(i.e., $\Delta A_0 = \Delta A(prolate) + \gamma_{\mathrm{in}}\int \phi\ {\mathrm {d}}A$),
the budding can be promoted  (see the green curve in Fig.~\ref{fig:bud}(b)).
The insertion can induce the budding even at $C_0=0$ through the protein binding to the large spherical region.

Lipid membranes supported on a solid substrate
are widely used as model systems for biological membranes,
providing a valuable platform to study both protein functions and membrane properties.\cite{tana05,cast06,acha10,ales14,weer15}
Boye and coworkers reported 
that the annexin proteins\cite{gerk05,bout15,blaz15} can detach lipid membranes from the substrate.\cite{boye17,boye18}
Their observation revealed membrane rolling and budding from open edges, with variations depending on the types of annexins.
The budding and vesicle formation observed in these experiments can be interpreted as the binding behavior of isotropic proteins.
Figure~\ref{fig:disk} shows the membrane detachment dynamics obtained by a meshless membrane simulation,\cite{nogu19c}
in which particles with a diameter of $\sigma$ self-assemble into one-layer sheets in a fluid phase.
The bound proteins (represented as red particles) induce membrane bending, counteracting the adhesion to the substrate,
leading to the formation of small vesicles from the membrane edge.

\begin{figure}[t]
\includegraphics[]{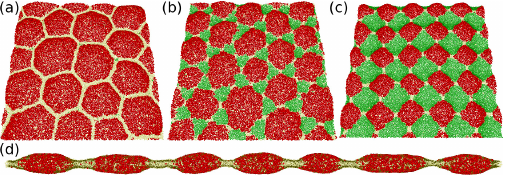}
\caption{
Phase separation induced by binding of isotropic proteins.
(a)--(c) Binding to upper and lower membrane surfaces
at $C_0\sigma= \pm 0.1$, $\kappa_{\mathrm{pi}}/k_{\mathrm{B}}T=144$, $\kappa_{\mathrm{d}}/k_{\mathrm{B}}T=16$, $\mu_{\mathrm{u}}/k_{\mathrm{B}}T =7.5$, and
$\mu_{\mathrm{ff}}=\mu_{\mathrm{d}}-\mu_{\mathrm{u}}$.\cite{nogu19c}
The red and green spheres represent membrane particles bound from the upper and lower surfaces, respectively.
The yellow spheres represent unbound membrane particles.
(a) Hexagonal pattern of the upper-bound domains in the unbound membrane at $\mu_{\mathrm{d}}/k_{\mathrm{B}}T = 4$.
Lower bound particles are negligible.
(b) Kagome-lattice pattern at $\mu_{\mathrm{d}}/k_{\mathrm{B}}T = 6$.
The upper- and lower-bound domains form hexagonal and triangular shapes, respectively.
(c) Checkerboard pattern at $\mu_{\mathrm{d}}=\mu_{\mathrm{u}}$.
Both upper- and lower-bound domains form square shapes.
(d) Beaded-necklace-shaped membrane tube induced by binding to the outer surface.\cite{nogu21b}
The red and yellow spheres represent bound and unbound membrane particles, respectively.
}
\label{fig:snapiso}
\end{figure}

\subsubsection{Phase Separation}\label{sec:psep}

Proteins exhibit both direct and membrane-mediated interactions, and
their assemblies often influence the membrane morphology.
The curvature generated by proteins can drive phase separation,
resulting in protein-rich and protein-poor membrane domains with distinct curvatures.
The vesicle budding process described in the Section~\ref{sec:bud}
represents an extreme case of phase separation, where protein-rich buds form in contrast to the protein-poor large spherical region.

Under conditions of high surface tension compared to the spontaneous curvature of the domain and the line tension of the domain boundary,
curved domains do not fully close into spherical buds  but instead adopt a spherical-cap shape.
When these spherical-cap domains expand to cover most of the membrane surface,
they organize into a hexagonal array, representing the closest packing configuration in 2D space, as shown in Fig.~\ref{fig:snapiso}(a).\cite{gout21}
As the binding chemical potential $\mu$ of proteins increases,
the membrane undergoes a continuous transition from an unbound state to a hexagonal phase.
This is followed by a first-order transition to the homogeneously bound phase,
where the entire membrane becomes saturated with proteins.\cite{gout21}

When proteins bind to both membrane surfaces from the upper and lower buffers,
the membrane can form both convex and concave domains, as shown in Fig.~\ref{fig:snapiso}(b) and (c).\cite{nogu19c,nogu25a}
Under symmetric conditions, where the chemical potentials of the upper and lower surfaces are equal ($\mu_{\mathrm{u}}=\mu_{\mathrm{d}}$),
the membrane exhibits distinct patterns depending on the chemical potential. 
At low chemical potentials,
square domains arranged in a checkerboard pattern obtained, 
while at higher chemical potentials, striped patterns emerge.
Small unbound membrane patches stabilize the vertices of the square domains (see Fig.~\ref{fig:snapiso}(c)).
When repulsive interactions are added between the unbound and bound membranes, these unbound patches expand and take on a square shape,
and the bound  domains adopt an octagonal shape, resembling the 4.8.8 tiling pattern.\cite{nogu25a}
Under asymmetric conditions, where the chemical potential of the upper surface exceeds that of the lower surface ($\mu_{\mathrm{u}}> \mu_{\mathrm{d}}$), 
a kagome-lattice pattern can form.
In this configuration, triangular concave domains are arranged within a hexagonal array of convex domains (see Fig.~\ref{fig:snapiso}(b)).
As the chemical-potential difference further increases, concave domains disappear and a hexagonal pattern of convex domains form (see Fig.~\ref{fig:snapiso}(a)).
Additionally, the transfer (flip--flop) of proteins between the two surfaces can be accounted for
using the flip--flop chemical potential $\mu_{\mathrm{ff}}$.
At thermal equilibrium ($\mu_{\mathrm{ff}}=\mu_{\mathrm{d}}-\mu_{\mathrm{u}}$),
the flip--flop does not change the equilibrium behavior owing to the principle of detailed balance.
However, under non-equilibrium conditions ($\mu_{\mathrm{ff}}\ne \mu_{\mathrm{d}}-\mu_{\mathrm{u}}$), the ballistic motion of biphasic domains 
and time-irreversible fluctuations of patterns can be observed.\cite{nogu25a}

Phase separation can also occur in both spherical and cylindrical membranes.
In spherical vesicles, the formation of hexagonal arrays of concave domains has been theoretically investigated.\cite{auth09}
In cylindrical membrane, a 1D periodic pattern can emerge, in which round bound and narrow straight unbound domains alternate in a beaded-necklace-like arrangement (see Fig.~\ref{fig:snapiso}(d)).\cite{nogu21b}

Even in the absence of spontaneous-curvature differences between bound and unbound membranes,
attraction between bound membrane regions can arise due to hydrophobic mismatch of transmembrane proteins and
Casimir-like interactions in rigid proteins. 
The height of the transmembrane proteins can differ from the thickness of surrounding membrane,\cite{lee03,ande07,vent05}
resulting in an effective attraction between proteins to reduce the hydrophobic mismatch.\cite{phil09,deme08,schm08,four99}
In thermal equilibrium,
the membrane height fluctuations follow the relation $\langle |h_q|^2 \rangle= k_{\mathrm{B}}T/(\gamma q^2 + \kappa q^4)$,
where $h_q$ represents the Fourier transform of the membrane height in the Monge representation.\cite{safr94,goet99}
Here, the surface tension $\gamma$ corresponds to the mechanical frame tension conjugated to the projected membrane area.\cite{shib16}
Rigid proteins with high bending rigidity $\kappa_{\mathrm{p}}$ suppress membrane fluctuations in their vicinity.
As a result, protein assembly mitigates entropy loss, leading to a Casimir-like attractive interaction.\cite{gout21,goul93}
This interaction is expressed in the leading order as $6k_{\mathrm{B}}T(r_{\mathrm{p}}/r)^4$,
where $r$ is the inter-protein distance and $r_{\mathrm{p}}$ represents the protein length.
Consequently, the binding of rigid proteins induces a first-order transition between unbound and bound states.\cite{gout21}
Additionally, Casimir-like interaction also arises between ligand--receptor pairs that connect adjacent membranes,
effectively reducing the fluctuations in the membrane separation distance.\cite{weil10,nogu13}

\begin{figure}[t]
\includegraphics[]{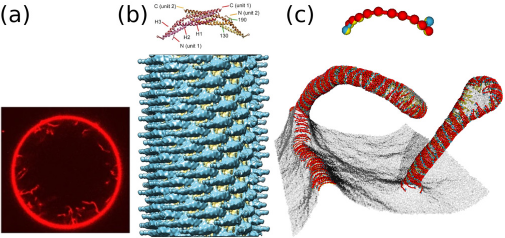}
\caption{
Tubulation generated by BAR domains.
(a) Confocal image of tubular invagination generated by the binding of I-BAR domains.
Reproduced from Ref.~\citenum{baro16}. Licensed under CC BY (2016).
(b) N-BAR (amphiphysin/B1N1s)-coated tube with a diameter of $280$\AA. 
3D reconstruction from cryo-EM images.
Reproduced from Ref.~\citenum{adam15}. Licensed under CC BY (2015).
(c) Tubulation simulated by a meshless membrane model.\cite{nogu19a}
The BAR domain and membrane beneath are modeled as a linear chain of (red and yellow) particles with two kink (light blue and yellow) particles for the molecular chirality.
In the upper panel, a protein rod is extracted to show the structure.
The gray spheres represent the bare membrane particles.
}
\label{fig:tub0}
\end{figure}

\subsection{Anisotropic Proteins}\label{sec:anigen}

\subsubsection{Interprotein Interactions}\label{sec:anipp}

For anisotropic proteins,
 excluded volume interactions are orientation-dependent, as discussed in Section~\ref{sec:aniden}.
Membrane-mediated interactions also depend on the protein orientation.\cite{nogu17,schw15,kohy19a}
In a tensionless membrane ($\gamma=0$),
 the curvature-mediated interaction energy for an isolated protein pair can be expressed in the leading order as\cite{nogu17}
\begin{eqnarray}
H_{\mathrm {pp}}^{(0)}(r_{12}) &=&
\frac{16 \pi {r_{\mathrm{p}}}^4}{9{r_{12}}^2}\kappa_{\mathrm {d}} C_{\mathrm {r1}} C_{\mathrm {r2}} \big[\cos(2\theta_1) \nonumber \\ 
&& +\cos(2\theta_2)-\cos(2\theta_1-2\theta_2)\big], \label{eq:Hint0}
\end{eqnarray}
where
 $\theta_1$ and $\theta_2$ denote the angles between the bending axis of proteins 1 and 2 and the vector $\mathbf{r}_{12}$
connecting their centers.
Two rigid proteins with curvatures $C_{\mathrm {r1}}$ and $C_{\mathrm {r2}}$ and a length of $r_{\mathrm {p}}$ are modeled as
 point-like objects\cite{nogu17,domm99,domm02,yolc14}, which allows the derivation of Eq.~(\ref{eq:Hint0}).
Similar angular-dependent interactions have been reported by assuming elliptical\cite{schw15} and circular protein shapes.\cite{kohy19a}

When two proteins bend the membrane in the same direction ($C_{\mathrm{r1}}C_{\mathrm{r2}}>0$),
they exhibit an attractive interaction when oriented side-by-side ($\theta_1=\theta_2=\pi/2$)
and a weaker repulsive interaction when aligned along the membrane axis ($\theta_1=0$ or $\theta_2=0$).
In the side-by-side dimer configuration (i.e., $\theta_1=\theta_2=\pi/2$), the membrane experiences reduced deformation.
This bending-energy reduction is the origin of this attraction.
Conversely, when the proteins bend the membrane in the opposite directions ($C_{\mathrm{r1}}C_{\mathrm{r2}}<0$),
 the interactions are reversed.
In this case,
the proteins exhibit weak attraction when aligned along the membrane axis ($\theta_1=0$ or $\theta_2=0$)
and repulsion when positioned side-by-side ($\theta_1=\theta_2=\pi/2$).
Therefore, proteins with similar curvatures preferentially interact in a side-by-side configuration,
whereas proteins with opposite curvatures prefer tip-to-tip alignment.
These interactions have been quantitatively confirmed through the meshless membrane simulations.\cite{nogu17}
Furthermore, the Casimir-like interaction  between straight rods exhibits a different angular dependence but
decay over a shorter range, proportional to ${r_{12}}^{-4}$.\cite{gole96,bitb11}

For positive surface tensions ($\gamma>0$), 
the bending energy dominates interactions on length scales shorter than $r_{\mathrm{ten}}=\sqrt{\kappa_{\mathrm{d}}/\gamma}$,
whereas surface tension effects become dominant at length scale greater than $r_{\mathrm{ten}}$.
As a result, the interaction energy changes from a bending-dominant regime to a tension-dominant regime at approximately $r_{12}\approx3r_{\mathrm{ten}}$:\cite{nogu17} 
\begin{eqnarray}
H_{\mathrm {pp}}(r_{12}) =&& \\ \nonumber
H_{\mathrm {pp}}^{(0)}(r_{12}),&& {\mathrm {\ for~}} r_{\mathrm{p}}<r_{12}\ll r_{\mathrm{ten}},
\\ \nonumber
H_{\mathrm {pp}}^{(1)}(r_{12}),&& {\mathrm {\ for~}} r_{12}\gg r_{\mathrm{ten}}{\mathrm {~if~}}\cos[2(\theta_1-\theta_2)]\ne 0,
\\ \nonumber
H_{\mathrm {pp}}^{(2)}(r_{12}),&& {\mathrm {\ for~}} r_{12}\gg r_{\mathrm{ten}}{\mathrm {~if~}} \cos[2(\theta_1-\theta_2)]=0,
\end{eqnarray}
where
\begin{eqnarray}
\frac{H_{\mathrm {pp}}^{(1)}(r_{12})}{\kappa_{\mathrm{d}}C_{\mathrm{r1}}C_{\mathrm{r2}}}&=&-\frac{64\pi {r_{\mathrm {p}}}^4{r_{\mathrm{ten}}}^2}{3{r_{12}}^4}
\cos[2(\theta_1-\theta_2)], \hspace{0.5cm}  \\ 
\frac{H_{\mathrm {pp}}^{(2)}(r_{12})}{\kappa_{\mathrm{d}}C_{\mathrm{r1}}C_{\mathrm{r2}}}&=&
\frac{(2\pi)^{3/2}{r_{\mathrm{p}}}^4}{9{r_{\mathrm{ten}}}^{3/2}{r_{12}}^{1/2}}\exp\big(-\frac{r_{12}}{r_{\mathrm{ten}}}\big)\Big[2 \\ \nonumber
&& +2\cos(2\theta_1)
 +2\cos(2\theta_2)+\cos(2\theta_1+2\theta_2)
\Big].
\end{eqnarray}

In some coarse-grained simulations,
the tip-to-tip assembly of crescent proteins on membranes has been reported.\cite{simu13,olin16}
In these systems, proteins sink into the bound membrane by a strong protein-membrane attraction, resulting in
 a strongly negative spontaneous curvature perpendicular to the major axis of the crescent proteins.
Consequently, the protein bending axis is perpendicular to the major axis, meaning that
tip-to-tip alignment, from the perspective of the protein's shape, corresponds to side-to-side alignment when viewed from the bending axis.\cite{nogu17}

\begin{figure}[t]
\includegraphics[]{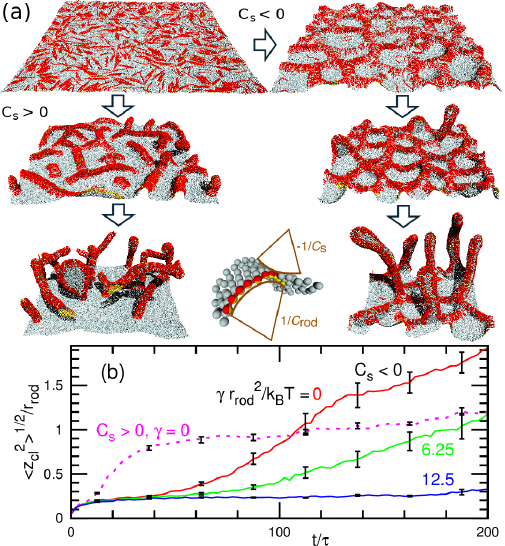}
\caption{
Tubulation from a flat membrane by crescent-rod proteins at $\phi_{\mathrm{rod}}=0.4$ and $C_{\mathrm{rod}}r_{\mathrm{rod}}= 4$.\cite{nogu16}
The proteins have the spontaneous curvatures $C_{\mathrm{rod}}$ and $C_{\mathrm{s}}$ along the protein axis and perpendicular (side) direction, respectively, as shown at the middle bottom in (a).
The initial state is an equilibrium state at $C_{\mathrm{rod}}= C_{\mathrm{s}} = 0$,
and the rod curvatures are tuned at $t=0$.
(a) The left panels show the sequential snapshots
at $t/\tau=0$, $12.5$, and $100$  for a positive side curvature ($C_{\mathrm{s}}r_{\mathrm{rod}} = 1$).
The right panels show the sequential snapshots at $t/\tau=10$, $100$, and $200$ 
for a negative side curvature ($C_{\mathrm{s}}r_{\mathrm{rod}} = -1$).
The chains of spheres (upper and lower half surfaces are in red and yellow, respectively) 
represent the protein rods, and
the gray spheres represent the bare membrane particles.
(b) Time evolution of mean cluster height $\langle z_{\mathrm{cl}}^2\rangle ^{1/2}$ normalized by the protein length $r_{\mathrm{rod}}$. The solid lines represent the data at the surface tension $\gamma r_{\mathrm{rod}}^2k_{\mathrm{B}}T=0$, $6.25$, and $12.5$ for $C_{\mathrm{s}}r_{\mathrm{rod}} = -1$. The dashed line represents the data at $\gamma=0$ for $C_{\mathrm{s}}r_{\mathrm{rod}} = 1$.
Reproduced from Ref.~\citenum{nogu16}. Licensed under CC BY (2016).
}
\label{fig:rflat}
\end{figure}

\subsubsection{Tubulation}\label{sec:tubule}

The binding of BAR superfamily proteins to the membrane induces the formation of tubules.
Tubulation from liposomes has been observed in {\textit{in vitro}} experiments (see Fig.~\ref{fig:tub0}(a)).\cite{itoh06,mim12a,fros09,shi15,baro16}
In living cells,
different types of BAR proteins localize to tubular membranes in specific organelles and membrane regions.\cite{mcma05,suet14}
Within these tubules, the helical assembly of BAR domains has been visualized using cryo-electron microscopy (EM), as shown in Fig.~\ref{fig:tub0}(b).\cite{mim12a,fros09,adam15}

\begin{figure}[t]
\includegraphics[]{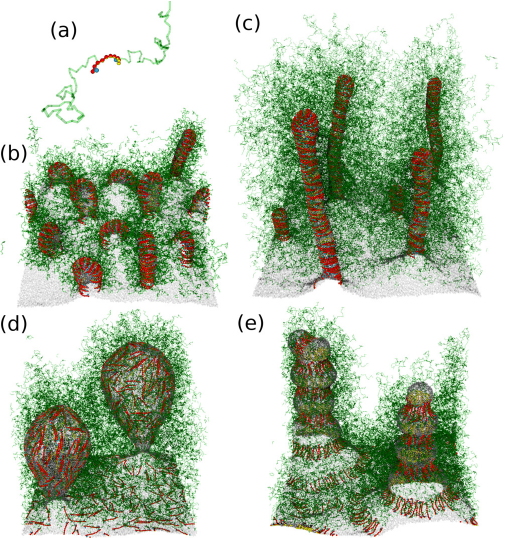}
\caption{
Tubulation and budding induced by crescent-rod proteins with anchoring excluded-volume chains at $\phi_{\mathrm{rod}}=0.24$.\cite{nogu22b}
(a) A protein comprising a crescent rod with two (light blue) kink particles (for chirality) and two excluded-volume chains of $n_{\mathrm{poly}}$ particles,
as a model of BAR proteins.
(b) An array of short tubules at  $C_{\mathrm{rod}}r_{\mathrm{rod}}= 3$ and $n_{\mathrm{poly}}=25$.
(c) Long tubules at $C_{\mathrm{rod}}r_{\mathrm{rod}}= 3$ and $n_{\mathrm{poly}}=100$.
(d) Ellipsoidal buds at $C_{\mathrm{rod}}= 0$ and $n_{\mathrm{poly}}=50$.
(e) Shish-kebab-shaped tubules at $C_{\mathrm{rod}}r_{\mathrm{rod}}= -3$ and $n_{\mathrm{poly}}=50$.
}
\label{fig:poly}
\end{figure}

\begin{figure}[t]
\includegraphics[]{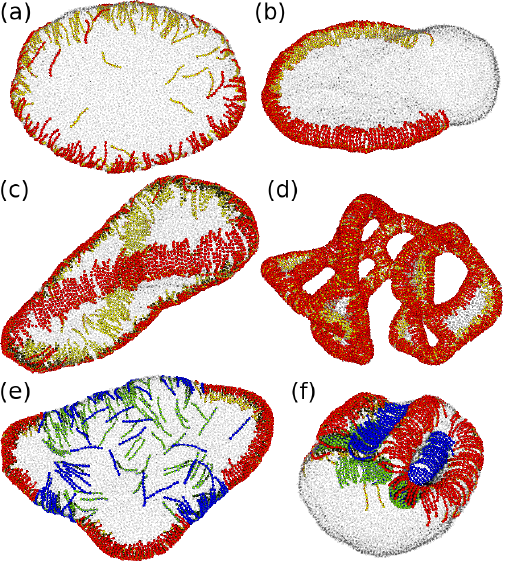}
\caption{
Snapshots of vesicles with crescent protein rods.
(a)--(d) A single type of protein is bound.
Here, a protein is represented by a linear chain of ten spheres, whose upper and lower halves are in red and yellow, respectively.
Unbound membrane particles are displayed in transparent gray.
(a) Disk-shaped vesicle at $\phi=0.167$ and $C_{\mathrm{rod}}r_{\mathrm{rod}}=2.5$.
The proteins are in the disk edge.\cite{nogu14,nogu15b}
(b) Linear protein assembly  at $\phi=0.167$ and $C_{\mathrm{rod}}r_{\mathrm{rod}}=3.75$.\cite{nogu14,nogu15b}
(c) Tetrahedral vesicle  at $\phi=0.4$ and $C_{\mathrm{rod}}r_{\mathrm{rod}}=2.5$.\cite{nogu15b}
(d) High-genus vesicle  at $\phi=0.8$ and $C_{\mathrm{rod}}r_{\mathrm{rod}}=4$.\cite{nogu16a}
(e)--(f) Two types of proteins are bound with the densities $\phi_1=\phi_2=0.15$.\cite{nogu17}
Two types of proteins are displayed in red and yellow and in blue and green, respectively.
(e)  Disk-shaped vesicle at  $C_{\mathrm{rod1}}r_{\mathrm{rod}}=4$ and  $C_{\mathrm{rod2}}r_{\mathrm{rod}}=2$.
The proteins are phase-separated in the disk edge.
(f) Vesicle with bumps  at  $C_{\mathrm{rod1}}r_{\mathrm{rod}}=3$ and  $C_{\mathrm{rod2}}r_{\mathrm{rod}}=-3$.
The linear protein assemblies with opposite curvatures are alternately aligned side-by-side.
}
\label{fig:snapsph}
\end{figure}

Tubulation and other membrane deformations have been realized using meshless membrane simulations
(Figs.~\ref{fig:tub0}--\ref{fig:snapsph}).
In these simulations, 
the protein rods are modeled as linear chains consisting of ten membrane particles, with or without two kink particles to account for chirality, as shown in Fig.~\ref{fig:tub0}(c).
The rod curvature  $C_{\mathrm{rod}}r_{\mathrm {rod}}\simeq 3$ corresponds to that of BAR-PH domains~\cite{zhu07},
where $r_{\mathrm{rod}}$ is the rod length.
Additionally, excluded polymer chains, each containing $n_{\mathrm{poly}}$ Kuhn segments to represent IDP domains, are incorporated, as shown in Fig.~\ref{fig:poly}(a).
Tubulation with a helical protein assembly can be effectively reproduced using meshless simulations of chiral protein rods (see Fig.~\ref{fig:tub0}(c)).\cite{nogu19a}
While tubulation can also be induced by the achiral protein rods, the chirality has been shown to enhance the tubulation process.\cite{nogu19a}

Figure~\ref{fig:rflat} shows the tubulation dynamics of the achiral straight crescent rods.\cite{nogu16}
The same type of protein rods exhibit a membrane-mediated attractive interaction when aligned side-by-side, as discussed in Section~\ref{sec:anipp}.
Consequently, these protein rods initially form linear assemblies perpendicular to their axis. 
Over time, the contacts of these assemblies lead to the development of a network structure at a sufficiently high protein densities.
Eventually, tubules protrude from the network (see Fig.~\ref{fig:rflat}(a)).
The stability of this network structure is influenced by the side curvature $C_{\mathrm{s}}$ of the proteins and the membrane surface tension $\gamma$.
A negative side curvature $C_{\mathrm{s}}$ reduces the bending energy at network branch points, leading to slower tubulation compared to the case where $C_{\mathrm{s}}>0$ (compare the dashed and solid lines at $\gamma=0$ in Fig.~\ref{fig:rflat}(b)).
Since tubulation results in a reduction of the projected membrane surface area, increasing membrane tension $\gamma$ inhibits tubulation (see three solid lines in Fig.~\ref{fig:rflat}(b)).\cite{nogu19a,shi15}

The addition of the IDP domains can either promote or suppress tubulation, depending on the conditions.\cite{nogu22b}
For a short IDP with $n_{\mathrm{poly}}=25$, the tubulation dynamics slow down and become trapped in a short-tubule array, 
as shown in Fig.~\ref{fig:poly}(b). In this case, the crowded IDP domains induce repulsion between tubules, preventing their fusion.
Conversely, when $n_{\mathrm{poly}}=100$, the IDP chains extend beyond the mean distance between tubules, allowing fusion 
and promoting tubule elongation in the vertical direction (see Fig.~\ref{fig:poly}(c)).
Thus, interactions between IDP chains and membranes enhance tubulation, 
while interactions between the IDP chains of neighboring tubules slow it down.
In the absence of spontaneous curvature in the binding domains, IDP domains facilitate the formation of ellipsoidal buds, since the IDP chains gain more conformational entropy in vertically elongated shapes (see Fig.~\ref{fig:poly}(d)). 
When IDPs are introduced to negatively bent crescent rods --where the binding domain and IDPs exhibit the opposite spontaneous curvatures--
periodically bumped tubules are formed (see Fig.~\ref{fig:poly}(e)).
For short IDP chains, the proteins assemble into a network structure, resembling Fig.~\ref{fig:rflat}(a), 
on the membrane. This assembly causes the membrane to become rugged due to the bumped assemblies.
Notably, a similar rugged vesicle has been observed in experiments involving a chimeric protein composed of I-BAR and IDP domains~\cite{snea19}.

\subsubsection{Other Membrane Deformations}\label{sec:defo}

Figure~\ref{fig:snapsph} shows vesicles deformed by the crescent protein rods.
In vesicles and membrane tubes, protein assembly occurs in two distinct steps as the rod curvature $C_{\mathrm{rod}}$ increases.\cite{nogu14,nogu15b}
At low $C_{\mathrm{rod}}$, the proteins are randomly distributed. As $C_{\mathrm{rod}}$ increases to an intermediate level, 
the vesicle deforms into a disk-like shape, with proteins concentrating at the disk edge (see Fig.~\ref{fig:snapsph}(a)).
At high $C_{\mathrm{rod}}$, proteins form an arc-shaped linear assembly, resulting in vesicle with flat disk and spherical regions (see Fig.~\ref{fig:snapsph}(b)).
In membrane tubes, proteins initially assemble in the azimuthal direction, causing the membrane to adopt an elliptic shape 
at a medium $C_{\mathrm{rod}}$. As $C_{\mathrm{rod}}$ further increases, proteins also assemble along the tube axis.
These assembly processes occur continuously, since each transformation progresses in a 1D manner.

At high $C_{\mathrm{rod}}$ and increasing protein density,
the length of the protein assembly exceeds the edge length of the disk-shaped vesicle.
Initially, the vesicle elongates into an elongated elliptical shape, eventually, forming polyhedral structures, such as a tetrahedral vesicle shown in Fig.~\ref{fig:snapsph}(c). 
In membrane tubes, this process results in polygonal deformations, with proteins assembling along the edge lines of the polygon vertices.\cite{nogu15b,nogu22a}
Unlike the continuous transition described earlier, the transformations between polyhedral vesicles and between polygonal tubes are discontinuous.\cite{nogu15b}
Notably, similar triangular membrane tubes have been observed in the inner mitochondrial  membranes of astrocytes.\cite{blin65,fern83}
At high $C_{\mathrm{rod}}$ and protein density, excessive protein-induce stress can lead to membrane rupture, giving rise to high-genus vesicles (see Fig.~\ref{fig:snapsph}(d)).\cite{nogu16a,ayto09}

When multiple types of proteins bind to a membrane, differences in their preferred curvatures can induce phase separation.\cite{nogu17,nogu17a,rama13,kuma22} 
When two types of proteins exhibit positive curvatures with different magnitudes, they can segregate into regions of large and small curvatures.
In the case of a disk-shaped vesicle, proteins with larger curvature preferentially assemble at the corners of the triangular disk (see  Fig.~\ref{fig:snapsph}(e)).
Conversely, when two types of proteins possess opposite curvatures, their 1D assemblies align alternately in a side-by-side
arrangement, forming periodic bumps (see Fig.~\ref{fig:snapsph}(f)).\cite{nogu17}
Within this alternating pattern, the different proteins establish tip-to-tip contact,
which is consistent with the attractive interactions in the tip-to-tip direction described in Section~\ref{sec:anipp}.
Notably, this alternating assembly can also occur in flat membranes; however, it is disassembled under high surface tension.\cite{nogu17}

Simulations showed that identical protein rods formed 1D linear assemblies through membrane-mediated interactions.
The introduction of direct inter-protein interactions can modify the assemblies.
The formation of helical tubular assemblies is further enhanced by direct attraction.\cite{nogu19a}
Specific types of direct interactions may be necessary to accurately describe the assemblies of certain proteins.
The endosomal sorting complex required for transport (ESCRT)
forms a distinctive assembly, characterized by
 a spiral-spring-like structure on flat membranes and a helical tube configuration on cylindrical membranes.\cite{juki23,woll09,mead22,azad23}
This spiral assembly is involved in endosomal fission.
In dynamically triangulated membrane simulations,\cite{rama12,rama13,rama18,kuma22} 
proteins are often represented as point-like inclusions with orientational degrees of freedom.
In their models, protein interactions are governed by an orientation-dependent yet laterally isotropic potential.
As a results, when the orientations and the distance between two proteins are fixed, the interaction energy remains identical for both side-by-side
and tip-to-tip alignments.
Owing to the attractive nature of this potential in both lateral directions, the resulting protein assemblies exhibit a thickness of a few proteins rather than forming a strict single-layer 1D structure.

\subsection{Adhesion of Nanoparticles}\label{sec:nano}

During phagocytosis, large objects, such as viruses and cell debris,
are engulfed by the plasma membrane and internalized into the cell.
The engulfment of colloidal nanoparticles has been extensively studied as a model system for phagocytosis, and 
nanoparticles are also widely considered as the carriers for drug delivery.\cite{behz17,dona19,alme21,mitc21}
Unlike curvature-inducing proteins, an adhesive spherical nanoparticle can become fully wrapped by the membrane;\cite{dasg17,nogu02a,li10,vand16}
however, as surface tension increases, the membrane undergoes a first-order transition to a partially wrapped state.\cite{dese03}
Similarly, liquid droplets can also be wrapped by the membrane, but in contrast, the partially wrapped droplets deform to satisfy the wetting conditions at the contact lines.\cite{dimo17,sata23}
For non-spherical particles, the wrapping process may be accompanied by changes in particle orientation.\cite{dasg17,dasg14,van24}

Nanoparticles exhibit membrane-mediated interactions, similar to those observed in membrane proteins.\cite{dasg17,reyn07,sari12}
Nanoparticles can induce the formation of membrane tubules, wrapping the nanoparticle assembly.\cite{sari12a}
Simulations of nanoparticles with crescent\cite{liu20} and hinge-like\cite{li24} shapes have been conducted as model systems for protein binding, 
revealing orientational assemblies analogous to those formed by  anisotropic proteins.
Note that these nanoparticles have negative spontaneous curvatures along their minor axes due to their rounded shapes.

\section{Summary and Outlook}\label{sec:sum}

This review examined the curvature-sensing and generation mechanisms of membrane proteins.
Laterally isotropic proteins are capable of sensing both the mean and Gaussian curvatures of membranes,
with their curvature dependence well described by the mean-field theory. 
The IDP chains increase the bending rigidity and spontaneous curvature of membranes, while
decreasing the saddle-splay modulus.
The binding of isotropic proteins can lead to the formation of spherical buds and periodic patterns, 
such as hexagonal, kagome-lattice, checkerboard arrangements, and beaded necklace tubes. 
The curvatures generated by proteins play a crucial role in stabilizing these phase-separated patterns.
Additionally, the insertion of hydrophobic segments can modify the area difference between the two leaflets of the bilayer within the ADE model,
ultimately inducing membrane budding.

The binding behavior of anisotropic proteins, such as those from the BAR superfamily proteins, depends not only on the membrane curvatures but also on protein orientations.
Orientation-dependent excluded-volume interactions can drive an isotropic-to-nematic transition among the proteins.
In the dilute limit, an isolated protein preferentially binds to wide cylindrical membrane tubes with its orientation 
aligned along the azimuthal or axial directions,
whereas it binds to narrow tubes with two distinct tilted orientations.
As protein density increases, these proteins undergo the first-order and second-order transitions from a state characterized by the coexistence of two tilt angles to an ordered phase with a single orientation angle, depending on the membrane curvature.

Anisotropic proteins are also capable of driving tubulation. Protein chirality enhances tubulation, 
whereas negative side curvature and positive surface tension counteract it.
The IDP domains of BAR proteins promote tubulation while simultaneously inhibiting tubule fusion,
leading to either accelerated or decelerated tubulation dynamics depending on the condition.
Furthermore, anisotropic proteins can facilitate the formation of disk-shaped and polyhedral vesicles, polygonal tubes, and periodically bumped membranes.

For a quantitative understanding of the curvature sensing and generation,
accurate estimation of protein bending properties is essential.
This review described the estimation of bending properties of I-BAR domains through curvature-sensing studies using tethered vesicles.
The same approach can be extended to other curvature-inducing proteins.
To analyze the effects of Gaussian curvature, comparisons between cylindrical and spherical membranes with equivalent mean curvature 
are particularly important, especially at large curvatures.
Additionally, the asymmetric protein shapes of proteins can be assessed by examining their orientation distributions in cylindrical and buckled membranes.
Molecular dynamics simulations of proteins on a buckled membrane\cite{gome16,mahm19} provide valuable insights into their curvature-sensing properties and behavior.

\begin{acknowledgments}
This work was supported by JSPS KAKENHI Grant Number JP24K06973.
\end{acknowledgments}


%

\end{document}